\documentclass[acmlarge]{acmart}

\usepackage[normalem]{ulem}
\newcommand{\revise}[2]{{\color{red}\sout{#1}}{\color{blue}#2}}
\newcommand{\newrevise}[2]{{\color{red}#2}}

\newcommand{\revthree}[1]{{\color{red}#1}} 

\renewcommand{\revise}[2]{#2} 
\renewcommand{\newrevise}[2]{#2}

\renewcommand{\revthree}[1]{#1}

\pdfoutput=1

\usepackage{graphicx}
\usepackage{multirow}

\usepackage{epstopdf}
\usepackage{siunitx}

\usepackage[textsize=tiny,colorinlistoftodos]{todonotes}
\makeatletter
\define@key{todonotes}{bh}[]{
	\setkeys{todonotes}{author=\textbf{Bin - notes}, color=lime!30}}%
\define@key{todonotes}{bh2}[]{
		\setkeys{todonotes}{author=\textbf{Bin - urges}, color=red!30}}%
\define@key{todonotes}{ak}[]{
	\setkeys{todonotes}{author=\textbf{Anthony}, color=blue!30}}%
\makeatother

\AtBeginDocument{%
  \providecommand\BibTeX{{%
    \normalfont B\kern-0.5em{\scshape i\kern-0.25em b}\kern-0.8em\TeX}}}

\setcopyright{acmcopyright}
\copyrightyear{2023}
\acmYear{2023}

\acmJournal{CSUR}
\acmVolume{XXX}
\acmNumber{XXX}
\acmArticle{XXX}
\acmMonth{1}

\usepackage{afterpage}
\usepackage{adjustbox}
\usepackage{capt-of}
\usepackage[printonlyused,withpage,nolist]{acronym}
\usepackage{tabularx,booktabs}

\begin{document}

\title{Chronicles of Jockeying in Queueing Systems}

\author{Anthony Kiggundu}
\affiliation{%
  \institution{German Research Center for Artificial Intelligence}
  \streetaddress{Trippstadter Strasse 122, 67663}
  \city{Kaiserslautern}
  \country{Germany}
}
\email{anthony.kiggundu@dfki.de}

\author{Bin Han}\authornote{B. Han is the corresponding author}
\affiliation{%
  \institution{RPTU Kaiserslautern-Landau}
  \streetaddress{Gottlieb-Daimler-Straße 47, 67663}
  \city{Kaiserslautern}
  \country{Germany}}
\email{bin.han@rptu.de}

\author{Dennis Krummacker}
\affiliation{%
  \institution{German Research Center for Artificial Intelligence}
  \city{Kaiserslautern}
  \country{Germany}
}
\email{dennis.krummacker@dfki.de}

\author{Hans D. Schotten}
\affiliation{%
 \institution{RPTU Kaiserslautern-Landau,}
 \institution{German Research Center for Artificial Intelligence}
 \city{Kaiserslautern}
 \country{Germany}
 }
 \email{hans_dieter.schotten@dfki.de; schotten@eit.uni-kl.de}
 
\renewcommand{\shortauthors}{Anthony Kiggundu, et al.}

\begin{abstract}

\revthree{Emerging trends in communication systems, such as network softwarization, functional disaggregation, and multi-access edge computing (MEC), are reshaping both the infrastructural landscape and the application ecosystem. These transformations introduce new challenges for packet transmission, task offloading, and resource allocation under stringent service-level requirements. A key factor in this context is queue impatience, where waiting entities alter their behavior in response to delay. While balking and reneging have been widely studied, this survey focuses on the less explored but operationally significant phenomenon of jockeying, i.e. the switching of jobs or users between queues. Although a substantial body of literature models jockeying behavior, the diversity of approaches raises questions about their practical applicability in dynamic, distributed environments such as 5G and Beyond. This chronicle reviews and classifies these studies with respect to their methodologies, modeling assumptions, and use cases, with particular emphasis on communication systems and MEC scenarios. We argue that forthcoming architectural transformations in next-generation networks will render many existing jockeying models inapplicable. By highlighting emerging paradigms such as MEC, network slicing, and network function virtualization, we identify open challenges, including state dissemination, migration cost, and stability, that undermine classical assumptions. We further outline design principles and research directions, emphasizing hybrid architectures and decentralized decision making as foundations for re-conceptualizing impatience in next-generation communication systems.
}
\end{abstract}



\received{09 January 2023}

\maketitle

\section{Introduction}

\revthree{Recent advances in information and communications technology (ICT) have spawned a class of applications that are sensitive to latency and to reliability. Examples include remote control and industrial automatio \cite{8840800,wollschlaeger2017future}, autonomous driving \cite{Yurtsever,Montanaro,garcia2020v2x}, and immersive XR services. These applications place unprecedented demands on packet transmission, task offloading, and real time resource allocation \cite{campbell2010autonomous}. These demands expose limits of classical queueing abstractions and motivate a fresh look at {\em impatient} queueing behavior, in particular {\em jockeying}: the act of switching jobs or users between service queues to obtain faster service.}

From the behavioral perspective, studies have categorized impatient consumers as those that observe queue status and refrain from joining, a manner termed as {\bf balking} \cite{Pazgal,8II}. Then there are those that join a queue and abandon it when the accumulated delay is more than expected {\bf(renege)} \cite{Stanford, Haight_renege}.  
\revthree{Less studied but operationally important is {\bf jockeying}, where a customer who has already joined a queue relocates to an alternative queue, often driven by perceived differences in waiting time, price or service capability \cite{kiggundu2024resource}.  Most analytical and experimental work on jockeying attaches costs, priorities or class labels to buffers to capture system heterogeneity and tenant preferences \cite{5}. This heterogeneity is the principal rationale for preferring one buffer over another, but it is not the sole trigger for switching.}

\subsection{Triggers and modeling variants}
Generally, the reasons that influence impatience among consumers as covered in most literature \cite{Mittal,STARK2019837} can be irrational with no consideration whatsoever for prevailing buffer conditions, or rational \cite{CARUELLE2023247}. \revthree{The following are the most widely studied triggers for impatience in queues:}

\paragraph{\revthree{Threshold-based switching:}} \revthree{ A common rule is a queue length or waiting time threshold; That is, when the difference in queue lengths exceeds a preset margin, an agent may move from the longer to the shorter queue \cite{Haight,Adan.Ivo.etal,Zhao1990TheSQ} or vice versa in some models \cite{wolfgangStadje}.  In some findings, jockeying is only permitted when the alternative queue is empty \cite{2,pezoa}. And in more versatile setups, entities can rationally choose to switch from any position within the queue to the end of an alternative queue \cite{10}. Or more aggressively intermingle randomly in what is referred to as pre-emptive jockeying \cite{He2013,harchol2010performance}.}

\paragraph{\revthree{Cost and expected delay criteria:}} \revthree{ In heterogeneous systems like mobile networks or cloud MEC platforms, simple queue length thresholds are often inadequate. Instead, tenants either premise this jockeying behavior on combined constraints on the queue length threshold and the expected waiting time \cite{EVANS1993897,11}. Other findings compare the expected remaining service time \cite{5,DOWN2006509} or the distance over which the workload must be migrated \cite{Bi0bjective,classica}. Or setups that consider the expected delay \cite{pezoa}. Essentially, the tenant’s decision to move around workload factors in information about changes in the system characteristics \cite{Hassin.Haviv,Clement}. This information in other findings is defined in terms of costs, subscription profiles or network traffic classifications \cite{Xingguo} to influence the rationality of the jockeying tenants \cite{wolfgangStadje}.
}

\subsection{Queue typology and baseline assumptions}\label{lbl:assuming}
Technically, queues are composed of multi-server or single server lines. The canonical \(A/S/c\) classification (where $A,S,c$
denoted the arrival rate, service interval and the number of \revthree{servers} available for processing respectively) introduced in the queueing literature \cite{KendallNotation} yields the familiar \(M/M/c\), \(G/G/c\), and \(E/PH/c\) families \cite{kleinrock1975queueing}. Table \ref{tab:queuing_notation} is a summary of commonly adopted queue notations symbolizing the statistical properties of arrivals, departures and service disciplines. 
\begin{table}[!hbtp]
     \centering
     \small
     \caption{Common queueing notation $A/B/X/Y/Z$, originated
from~\cite{KendallNotation}}
     \label{tab:queuing_notation}
     \begin{tabular}{llll}
         \toprule[2px]
         Characteristic            &&Symbol &Explanation and
remarks\\\midrule[1.5px]
         \multirow{7}{5cm}{Inter-arrival interval ($A$)\\Inter-service
interval ($B$)}
&
                                  &$M^{[S]}$ &Exponential with random
batch size $S$,\\
                                  && &$S\equiv 1$ by default\\
                                 &&$D$ &Deterministic\\
                                 &&$E_k$ &Erlang type $k$ ($k=1,2,\dots$)\\
                                 &&$H_k$ &Mixture of $k$ exponentials\\
                                 &&$PH$ &Phase type\\
                                 &&$G$ &General\\\hline
         Number of servers ($X$) &&\multirow{2}{*}{$1,2,\dots,\infty$}
&\multirow{2}{*}{$Y=\infty$ by default}\\
         Maximal queue capacity ($Y$) &&&\\\hline
                                 &&FCFS &First come, first served
(default) \cite{Brandwajn,Nicol} \\
         \multirow{2}{*}{Queue discipline ($Z$)}
                                 &&LCFS &Last come, first served \cite{70296,10.2307/2667250}\\
                                 &&RSS &Random selection for service \cite{Kingman1962OnQI}\\
                                 &&PR &Priority \cite{XieHeZhao,Takagi1991,Andreas.Manfred}\\
                                 &&GD &General discipline\\
         \bottomrule[2px]
     \end{tabular}
\end{table}
\revthree{Several standard assumptions recur across jockeying studies in addition to the above nomenclature:}

\begin{itemize}\setlength\itemsep{0pt}

  \item \revthree{ {\bf Arrival and service processes.}  Although studies that differentiate between the properties of the statistical distributions of both arrivals and departures in buffers exist \cite{Ferng,Collings}, most findings generalize the periodicity of these two activities to obey a Poisson distribution \cite{poisson1837recherches,departs}. Others presume this frequency as batch arrivals \cite{Zadeh2015,choudhury2004two} for example in systems with limited capacity \cite{Takagi1993Part2}, and the performance effect that the periodicity of these two events has on the impatiently queueing consumer has been investigated in \cite{kleinrock1976queueing,harchol2010performance,bin_slaas}. The dynamics introduced by these events in queues are mostly generalized to form continuous or discrete stochastic chains \cite{Takagi1993Part3}.}
  \item \revthree{{\bf Service discipline.} First-Come-First-Served is the default service discipline in most work, although according to Table \ref{tab:queuing_notation} other disciplines have been documented \cite{kleinrock1975queueing}. However, practitioners argue that mission critical operations, such as rescue (where tasks urgently need to be prioritized by jockeying) render the FCFS approach less feasible given the strict latency requirements in such operations \cite{Takagi1991,Kogan1997,Kumar2021}.}
  \item \revthree{{\bf Homogeneous setups.} Another wild assumption in most models is that the setups are homogeneous in nature, further limiting the applicability of some modeling techniques in next generation communication systems like 5G and Beyond. } 
  \item \revthree{{\bf Join the Shorter Queue (JSQ).} It has also been argued that joining the shorter queue might not be optimal given the existential differences in queue capacities and workload \cite{shorter_queue}. Limiting jockeying to occur from the shorter buffer to the longer one raises concerns about such jockeying thresholds as the optimal trigger for switching buffers in heterogeneous setups.}
  \item \revthree{{\bf Queue length difference as a jockeying threshold.} The use of a preset difference in the size of the buffers (jockeying threshold) as the criterion for the task migration is not practical in heterogeneous systems.}
  \item \revthree{{\bf Information availability.} A common modeling premise is that arriving agents have access to some form of queue status information \cite{Jeongsim}. This information guides the decision to join a queue and the common strategy is to join the shorter one \cite{I.J.B.F,ravid_2021}, or switch to an alternative queue \cite{Nicol,Melamed}. Technical mechanisms for providing tenants with access to this buffer status information have inspired proposals for broadcasting or subscription to this information in dedicated or shared communication channels \cite{7}.  
  } 
\end{itemize}

\subsection{\revthree{Why classical assumptions may fail in MEC or 5G}}
\revthree{Emerging architectural trends like network softwarization, functional splits in the RAN ( Radio, Distributed, Centralized Units and Control or User Planes), network slicing and multi-access edge computing (MEC) are fundamentally altering the latency, control plane and orchestration semantics that underlie modern communication systems \cite{JiangBin}. These transformations introduce tight service-level constraints, multi-domain heterogeneity and new costs (e.g., state transfer latency, orchestration handshakes, and control plane signaling) that violate key assumptions in many classical impatient queueing models.}

\revthree{In particular, common primitives such as Markov modulated queues, threshold-based jockeying rules, Erlang/C waiting-time approximations, implicitly assume fast and reliable dissemination of queue descriptors, negligible migration cost, stationarity and trusted status reports. Under \ac{O-RAN} proposals for disaggregation and multi-vendor slicing, these assumptions break down. This is because information is partial and delayed, migration incurs nontrivial transfer and reconfiguration overhead, and decision policies face incentive and security constraints across administrative domains.
Therefore, while the modeling assumptions highlighted in Section \ref{lbl:assuming} are useful, the following aspects of modern MEC and 5G systems undermine their validity:
}
\begin{enumerate}\setlength\itemsep{0pt}
  \item \revthree{{\bf Heterogeneity and multi-vendor slices.} Slices assemble diverse RAN, core and MEC components with distinct performance and cost profiles \cite{barakabitze20205g,Elayoubi,Ameigeiras}. A simple queue length comparison does not capture differences in processing power, transport latency, or the monetary costs that tenants incur when switching between slices \cite{Papageorgiou,Sciancalepore,shu2020novel,wang2019enable}.}
  \item \revthree{{\bf Non-zero migration cost.} Stateful migration of tasks or sessions incurs transfer time proportional to state size and network path characteristics\cite{GAETA2006149,Maekawa}. Therefore, ignoring transfer cost biases models toward excessive switching.} 
  \item \revthree{{\bf Information latency and overhead.} Timely and authenticated dissemination of per-slice descriptors (waiting time, load, price) consumes control plane resources\cite{Hassin.Haviv}. Yet stale or noisy descriptor information could yield poor decisions and may increase churn \cite{ImpatienceBin,Tolosana,Longbo,Shiang_etal}.}
  \item \revthree{{\bf Stability and security risks.} Low queue length based jockeying thresholds can induce a "ping-pong" behavior or oscillations that could amplify the load \cite{Hassin.Haviv,Gertsbakh1984TheSQ}. Furthermore, unauthenticated status reports can be exploited for selfish or adversarial gain, necessitating trust mechanisms \cite{Kholidy,Schinianakis}.} 
\end{enumerate}

\subsection{\revthree{Implications and objectives of this survey}}
\revthree{Proposed Sixth Generation (6G) architectural modifications by the \ac{3GPP} fundamentally challenge the assumptions underpinning classical jockeying models. Partial and delayed visibility, state transfer costs, heterogeneous slice performance, and cross-domain trust constraints jointly undermine the tractability of simple threshold or Markovian-based decision rules. In particular, \textit{heterogeneity and multi-vendor slicing} introduce non-uniform queue capacities, service disciplines, and admission rules, rendering traditional homogeneous formulations inapplicable. At the same time, \textit{information latency and dissemination overhead} become critical in MEC and \ac{SDN}/ \ac{NFV} deployments, where stale or excessive signaling can negate the benefits of jockeying decisions. Moreover, \textit{stability and security risks}, including oscillatory switching (ping-pong effects) and adversarial or corrupted state reports, threaten SLA compliance and trust in decentralized decision making. Taken together, these transformations render many existing jockeying models infeasible, and they highlight the pressing need for next-generation approaches that explicitly incorporate slice heterogeneity, communication-constrained information flows, and robustness guarantees under adversarial or unstable dynamics.}

\revthree{Motivated by these shifts, this survey consolidates the state of the art in jockeying models for communication systems, identifies the limitations of existing techniques under emerging 5G/6G architectural paradigms, and outlines the critical directions where future modeling and system design efforts must focus. Hence, the contributions of this chronicle can be summarized as follows: }
\begin{enumerate}
  \item \revthree{\textbf{Taxonomy and synthesis.} This chronicle catalogs existing jockeying models and control primitives, highlighting their assumptions, analytical tools, and applicability. It provides a retrospective survey of techniques for modeling this impatience behavior, for which we know no equivalent compilation has been published.
  } 
  \item \revthree{\textbf{Quantitative gap analysis:} We evaluate the \emph{practicality} of jockeying under contemporary architectural constraints. That is, we assess the limits of classical models in 5G/6G contexts, focusing on heterogeneity, signaling constraints, and security stability challenges.}
 \item \revthree{\textbf{Integration with emerging architectures:} We analyze how \ac{MEC}, \ac{SDN}/\ac{NFV}, and network slicing redefines modeling jockeying behavior by introducing heterogeneity, signaling delays, and cross-domain trust constraints. This unravels  where classical homogeneous and fully observable impatience modeling no longer applies.}    
  \item \revthree{ \textbf{Design criteria and Future directions:} From the taxonomy and gap analysis, we outline design principles and research directions for developing robust, communication-aware, and jockeying models suitable for next generation distributed environments. These preeminent challenges include but are not limited to migration latency and bandwidth cost, dissemination delay and staleness, multi-slice heterogeneity, robustness to stale or noisy information, prototype validation on MEC or \ac{NTN} testbeds etc.
  }
\end{enumerate}

In Section II of this chronicle, we provide a concise overview of the principal techniques used to model jockeying in queues. Following that, we classify the literature by methodological family and examine individual studies in each class, emphasizing their models, assumptions, and numerical or empirical results. We intentionally avoid lengthy formal proofs; where analytical results are essential to the argument, we summarize the key theorems or lemmas and state their implications for jockeying policies. In the discussion section we then assess how recent architectural proposals (notably those advocated by the O-RAN community) affect the practicality of existing jockeying models and policies. Motivated by that assessment, we identify concrete open problems and promising research directions. Particularly, those that address dissemination overhead, task migration costs, performance stability, security, and the deployment challenges of impatience aware mechanisms in MEC and slice-based deployments.
\revthree{The following sections systematically review existing jockeying techniques before analyzing their applicability to next-generation networks.}

\section{\revise{}{Techniques for Modeling Jockeying in Queues}} 
\revise{}{
Different approaches have been studied to characterize for equilibrium conditions and optimize queue descriptors for performance enhancement. \newrevise{These approaches model the dynamics associated with impatient queueing.}{ Table \ref{table:sum} provides a comparative summary about the individual models grouped based on the similarities in techniques while differentiated along performance and descriptor evaluations.}
}

\subsection{Stochastic Modeling}
Most literature models jockeying as a Markov decision problem: \ac{MDPs} (and their generalizations \ac{POMDPs} or \ac{Dec-POMDPs}) formalize sequential decision-making under uncertainty and are natural for planning agent actions in dynamic environments \cite{natarajan2022planning,Monahan,Tahir,Alsheikh,Jiekai,Pajarinen}. However, these approaches suffer well-known practical limitations: the exploration–exploitation tradeoff can trap learning in local optima, state-space growth renders exact policy search computationally intractable for realistic multi-agent settings, and partial observability further complicates inference and control \cite{goldsmith1998complexity,Scherer}. In short, despite strong theoretical grounding, MDP-based methods often struggle to scale and to provide timely, high-quality decisions in large, heterogeneous MEC/NTN deployments.

Game-theoretic formulations (Nash equilibrium and variants) offer an alternative by characterizing selfish participants’ strategies and equilibrium outcomes \cite{Nash,Refael_Hassin,Hassin.Haviv,Laan2021,Haviv2015StrategicBI}. Yet they too rely on demanding assumptions like players’ knowing the strategies of others players and can be slow to converge\cite{Gatti,Daskalakis}. These techniques are also sensitive to initialization, or susceptibile to saddle-point dynamics in sequential or repeated interactions\cite{Samid,JORDAN1993368,Zaiming} . Consequently, while both MDP/POMDP and Nash-frameworks yield important insights, their computational and information assumptions limit direct applicability in large, decentralized, latency-sensitive networked systems.

\newpage
\clearpage
 \begin{table*}
   \caption{Different approaches used for analyzing the jockeying behavior of customers in queueing systems}
   \begin{adjustbox}{width=\textwidth}
   \begin{tabular}{|p{8.3cm}|p{1.5cm} |p{2.5cm} |p{5.5cm}|}

   \toprule
         & \textbf{Queue model} & \textbf{Jockeying threshold}  & \textbf{Other metrics} \\ \midrule
 \multicolumn{4}{@{}l}{\textbf{Stochastic jockeying models}}\\
    \hline
    A Solution for Queues with Instantaneous Jockeying and Other Customer Selection Rules &  M/M/C   &   2  &  Distance traversed by jockeys    \\
    On Jockeying in Queues &   M/M/2   & 1  &  mean waiting times, impact of jockeying on service rate   \\
    Two queues in parallel &   M/M/2  &  1  & mean queue sizes \\
    The shortest queue model with jockeying &   M/M/C   & 1 & mean queue sizes, mean waiting times \\
    General solutions of the jockeying problem &    M/M/3 + M/M/$\infty$   & $\ge 2$    & effect of arrival rate on jockeying rate, system utilization  \\
    Dynamic routing and jockeying controls in a two-station queuing system &  M/M/2  & $\ge 1$  & Workload distribution, cost efficiency \\
    Optimal policy for controlling the two-server queueing systems with jockeying & M/M/2  & Hedge point & Cost optimization \\
    Dynamic Load Balancing in Parallel Queuing Systems &  M/M/2 + M/M/C   &  $\ge N$ & frequency of monitoring queues \\    
    Strategic Dynamic Jockeying Between Two Parallel Queues &  M/M/2  &   &  queue lengths limits for jockeying \\
    A new look on the shortest queue system with jockeying &  M/M/C  & $\ge N$  &  Count of jumps before being served \\
    
    A Queuing System with Two Parallel Lines, Cost-Conscious Customers, and Jockeying &  M/M/2   &  $\ge 1$   & Costs optimization  \\
    Analysis of job transfer policies in systems with unreliable servers &     &   $\ge N$  & Workload expiry threshold  \\
    Stability analysis of some networks with interacting servers &     &  $\ge N$   &    \\
    Two M/M/1 Queues with Transfers of Customers &  2xM/M/1  &  1   &  transfer rates, time to migrate workload, magnitude of bulk transfer  \\
    Tail asymptotics of two parallel queues with transfers &  M/M/2   &   $\ge N$  &    \\
     Energy-efficient heuristics for job assignment in server farms &  M/M/2 &  $ $ & Energy Efficiency  \\
  \midrule
     \multicolumn{4}{@{}l}{\textbf{Analytic Models}}\\
     \hline
    Matrix-geometric analysis of the shortest queue problem with threshold jockeying &  M/M/C & $\ge N$ & mean queue size, mean waiting time   \\
    Analysis of two queues in parallel with jockeying and restricted capacities &  M/M/2 + 2xM/M/1 & iff N1=0 or N2=0 & traffic intensity \\
    A matrix-geometric solution of the jockeying problem & M/M/C  & $\ge 2$  & mean queue sizes, mean waiting time \\
    The shorter queue problem: A numerical study using the matrix-geometric solution &  M/M/2   & $10<n>\infty$ & mean queue length, mean waiting time  \\
    
    Queuing Analysis of a Jockeying & M/M/C  & $\ge 1$ & waiting time, traffic intensity \\
    Transient analysis of two queues in parallel with jockeying &  M/M/2   &  iff N1=0 or N2=0  & traffic intensity \\
    Analysis of the asymmetric shortest queue problem with threshold jockeying &  M/M/2 &  $\ge N$ & mean queue length  \\

   \midrule
   \multicolumn{4}{@{}l}{\textbf{Behavioral models}}\\
   \hline
    Dynamic load balancing in distributed systems (regeneration-theory) &   M/M/2  &  Delay &  mean task completion time \\
    Bi-objective optimization for multi-layer location–allocation with jockeying &   M/M/C  &  Fitness function &  Idle time, transfer overhead  \\
    Equilibrium strategies and value of information in two-line systems &  M/M/2 &  $3<n>\infty$ & queue status information cost \\
    Impatient queuing for intelligent task offloading in MEC &  G/G/1/$\infty$   &   1  & Costs, response time and rewards  \\
    Resource Allocation in Mobile Networks: A decision model of jockeying in queues &   M/M/C  &  Expected waiting time &  jockeying frequency, waiting time  \\
    \revthree{Information Bulletin Strategy in Impatient Queueing} &   \revthree{M/M/2}  &  \revthree{ Markov information models} &  \revthree{impatience rates, waiting time, service optimality } \\
  \bottomrule       
   \end{tabular}
   \end{adjustbox}  
   \label{table:sum}
  \end{table*}
 \clearpage

Analogous to continuous fluid flow, Fluid models are widely used for infinite-capacity queueing approximations and for studying bursty traffic regimes \cite{Vandergraft,Elwalid1991,GAETA2006149}. They have been applied to evaluate loss rates, maximum buffer sizes, busy periods and asymptotic/equilibrium behavior, sometimes using matrix-geometric or diffusion (Brownian) variants \cite{Soohan,Rajeeva}. Extensions incorporate feedback of queue-length information so transition rates depend on buffer state \cite{SCHEINHARDT2005551,Mehmet}. However, fluid descriptions are inherently nonstationary for heterogeneous traffic, produce nonlinear transient equations that are hard to solve, and demand substantial computation for sensitivity analysis, optimization and validation \cite{Lassila2014}.
        
\subsection{Analytic Modeling}
Matrix-geometric methods provide tractable steady-state solutions for Quasi-Birth-Death and other infinite Markov chains by exploiting the generator’s repeating block structure \cite{neuts_1986,Mitrani}. The idea is to essentially identify the irregular (initial) and regular (repeating) portions of the representative generator matrices that encode the various states a system (re)visits \cite{Haverkort,KAO199067}. Under mean-drift conditions one decomposes the generator into initial and repeating blocks and computes the rate (R) matrix whose spectral structure yields equilibrium probabilities \cite{Kapodistria}. While powerful, there are concerns about using cyclic or logarithmic reduction techniques to solve the rate matrix being computationally complex. Also, the solution for this rate matrix depends on the partitioning scheme between the state sub-levels to compose for initial and boundary states, making the unique solution more intractable. 

\subsection{Behavioral Modeling}
\revise{A promising new paradigm, the methodology is motivated by propositions for introducing decentralized control
of the impatience behavior in queueing systems. Here, the modeling deviates from canonical techniques that
assume centralized control of this behavior [ 55, 82]. An example in this class of models are artificial neural
networks, which are connectionist approaches that posit mathematical encapsulation of cogitative abilities.
Neural networks define for ways to obviate the high dimensionality curse suffered by statistical modeling of
complex systems. The networks can implicitly identify complex nonlinear relationships between dependent and
independent variables. Hence their deployment in the quantitative prediction of queueing descriptors [11, 85 ].
Encapsulating complexity in a “black box” however makes it difficult to understand how the predictions are
made. Other disadvantages of these approaches are the computational burden, proneness to under or over-fitting,
immature convergence, hyper-parameterization [16, 163], stopping rules, etc [104, 126].}{A growing alternative is decentralized, data-driven behavioral modeling that treats impatience and jockeying as local decision problems rather than centrally controlled processes \cite{7,kiggundu2024resource}. Artificial neural networks and related connectionist methods can capture high-dimensional, nonlinear relations and predict queue descriptors where classical models fail \cite{Minsu_etal,Aussem_etal}. Their drawbacks are well known: opacity (“black box”), training costs, over/under-fitting risks, hyperparameter tuning, convergence/stopping issues and additional compute burden at inference time \cite{yang2020hyperparameter,bergstra2011algorithms,Lodwich,PRECHELT1998761}
}

\revise{}{\section{Stochastic Models}}
\subsection{\revise{}{Statistical Models}}
\newrevise{Preliminary findings to model for}{Findings that investigate} the behaviour of impatient consumers when queueing up for resources can be traced back to a paper by \cite{Haight}\newrevise{, in which a  simple setup with only}{. There simple setup was composed of} two service lines, one ``near'' and the other ``far''. Whether ``near'' or ``far'' queue was defined by the queue sizes $X(t), X^{'}(t)$ at the time $t$ of an admission respectively. Such that, the inequality $X(t) \leq X^{'}(t)$ meant new customers preferred to join the shorter buffer (``near'') line\newrevise{and}{. On the other hand, existing consumers} were allowed to switch to the ``far'' queue when deemed beneficial.
\newrevise{The author analyzed cases two of customers arriving to stations following a Poisson distribution at rate $\lambda$. The customer had the choice to stay in a given queue operating at Poisson distributed service rate $\mu_i$ ($i=1,2$) or jockey to another}{New customers arrived to the ``near'' or ``far'' queue following a Poisson distributed with rate $\lambda$; And were processed at Poisson distributed service rates $\mu_{1}$ and $\mu_{2}$ respectively}. The switching happened when the queue sizes varied by one and this was always from the longer to the shorter line. The objective was to formulate expressions for steady state conditions in infinite time.
\newrevise{The authors first considered for the scenario when no jockeying was allowed such that for each change in state given any action (like a new arrival, exit or both actions happening concurrently in either service lines at a given time), expressions for the rate of change in the queue sizes $\frac{\delta p{xy}(t)}{\delta t}$ were first derived. The formulation then expressed for $p_{xy}(t)$ as $t\rightarrow\infty$ ({\itshape where $p_{xy}(t)=Pr\{X(t)=x,X^{'}(t)=y\}$}) to yield the bi-variate generating representation}{Expressions for the rate of change in queue sizes $\frac{\delta p{xy}(t)}{\delta t}$ were first derived. Then, from the formulations of the queue occupancies $p_{xy}(t)$ as $t\rightarrow\infty$ ({\itshape where $p_{xy}(t)=Pr\{X(t)=x,X^{'}(t)=y\}$}) evolved Equation~\eqref{eq:1} as the bi-variate generating function that defined for equilibrium conditions.} \newrevise{as a product over state space changes $s^{x}s^{'y}$ and a subsequent summation over this product to denote for the generating function Eq.~\eqref{eq:1}}{This was a product over state space changes $s^{x}s^{'y}$ whose summations characterized for the generating function}. Partial derivations of this generating function were then evaluated under variations in queue states $[(s=s^{'}=0), (s=1,s^{'}=0), (s=0, s^{'}=1), (s=s^{'}=1)]$ \newrevise{for each of these activities.}{given arrivals, exits or both events happening simultaneously:}
\begin{equation}
    \Phi(s,s^{'})=\sum_{x}\sum_{y}P_{xy}s^{x}s^{'y},
    \label{eq:1}
\end{equation}\textit{where $s$ and $s^{'}$ denoted the states of the ``near'' and ``far'' queues in terms of their queue lengths respectively.}

The distributions of the sizes for each queue and the overall system occupancy were formulated for from the representational difference equations.  
\newrevise{The paper concluded with the proof}{In conclusion, the findings derived} for the stability conditions of the queue lengths when customers needed to switch from one line to another. And the solutions for these conditions in each of the states were expressed for in terms of probability that both queues were not occupied. 

\begin{figure} 
 \centering
  \includegraphics[width=0.42\textwidth]{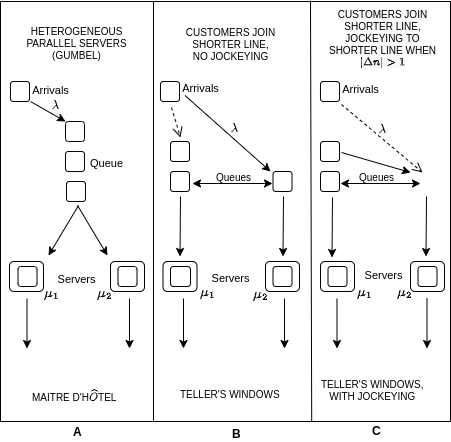}
  \caption{\scriptsize Jockeying strategies: In the left-most (A) of the illustration is the Maitre d'Hotel queueing strategy where customers waited in a single line and got served when one of the available stations was empty. In the middle (B) is the Tellers' Window strategy where customers joined and waited in the shorter of the two queues and no switching lines was permitted thereafter. In the right-most (C) was the Tellers' Window with Jockeying, a behaviour where despite a new customer having joined the shorter of the two queues, switching to an alternative one was permitted later given a deviation in the sizes by one.}
  \Description[Examples of queuing setups]{In extreme left setup orchestrated arriving customers accessing one of the two empty queues but not allowed to switch from one to another. Next to this was another setup where waiting customers joined the queue with the least number of consumers and on the extreme right was a setup where switching from a longer to a shorter line.}
   \label{fig:maitre_koennigberg}
\end{figure}

\cite{10} would later on conduct more concrete studies that compared heterogeneous queue setups\newrevise{and rules where}{. The findings assumed} tenants could instantaneously jockey given some threshold on queue length difference (adjacent queue shorter by one) or jockey based on some stochastic computations (based on how the differences in sizes of the queue changed).
Figure \ref{fig:maitre_koennigberg} was a depiction of the heterogeneity in setup (also referred to as strategies). Here, each setup was associated with a set of behavioral rules for the customers.
\newrevise{}{For example,} under the "Tellers' Windows with Jockeying" strategy (Figure \ref{fig:maitre_koennigberg}C), new arrivals were queued to the end of the shorter line. \newrevise{and could}{They could however} move to the other queue when the difference between the two queues exceeded one i.e. instantaneous strategy. \newrevise{ or based on the rate ($k(w_{i}-w_{j})$) {\itshape ($w_{i},w_{j}$ as queue sizes)} at which the queues varied in size (probability strategy)}{The case of the stochastic-based jockeying strategy on the other was a factor of the rate $k(w_{i}-w_{j})$ {(\itshape where $w_{i},w_{j}$ as queue sizes)} at which the queues varied in size.}
Specific of the probabilistic jockeying, the structure of the generator function and the evaluations for the steady state conditions \newrevise{revealed that the behavior of the system was the same as the setup}{exhibited known properties. These properties were descriptive of a phenomenon observed in similar setups } where customers simply joined the shortest queue and never left until the end of service ("Teller's Window"). 
For the instantaneous jockeying, \newrevise{equations were derived}{equilibrium conditions were characterized for} under the three presumed occupancy state categories $(n,n),(n+1,n)$ and $(n,n+1)$ ({\itshape where $n$ was the average number of occupants in a given queue})\newrevise{}{. These categories restricted jockeying within the system to the shorter line always} .
More complexity was introduced in the "Lane Changing" strategy (setup) where switching was not based on a preset jockeying threshold but the rate at which the two service lines differed in length. It was however observed that under this setup, the number of customers that wanted to jockey grew exponentially and that equilibrium conditions were a factor of only $\lambda, (\lambda=\lambda_{1}+\lambda_{2})$ and not $\lambda_{i}, i=1,2$. In addition to that, it was noted that some of these strategies led to states where one service line was empty while the other had customers queueing up, states in which queue utilization was compromised. 
Another interesting customer flow "Route changing" was presented where jockeying was probabilistic, depending on how dissatisfied a customer was being at a certain position in the queue. This was given the fact that customer had no access to queue status information \newrevise{, and the}{. The probability that a customer moved from queue $i$ to $j$ at time $t$ was therein computed as
$
P_{\overrightarrow{ij}}(t) = \begin{cases}
0,   & n_{i}=0 \\
 1-e^{-k_{i}(n_{i}-1)t}, & n_{i} \ge 1    
\end{cases}
$
}%
\textit{where $n_{i}$ \revise{}{was} the number of users in queue $i$ and $j$ was the preferred queue}. 

\noindent \newrevise{It was proven that expressions }{From the transition equations, the following system descriptors were characterized for: }the equilibrium conditions of the average queue occupancy, the expected number of customers processed and the probability that the a queue was not occupied. The work finally presented results from the numerical studies that involved experimenting with varying queue parameters under the aforementioned strategies (setups)\newrevise{ and shared insights into}{. Further insight was shared about} the queues' performance and quantitative measures on certain queue descriptors.

\newrevise{It would later be revealed that}{It would later be revealed that, using the generating function to formulate for the steady-state solution was not the only approach to the problem.} \newrevise{showed that}{According to }\cite{classica}, a closed form solution for some queue descriptors like queue length could be derived from inversion operations on the representational state space matrix. 
Customers that joined the $M/M/\revise{}{C}$ queue system were governed by different rules at admission (like which queue to join based on probabilities or queue size). \newrevise{in cases of switching queues (say $n_{1}-n_{2}\ge2$)}{Jockeying  was permitted only when the difference between the sizes of the two queues hit the preset threshold (two)}.
The authors differed in methodology to reason that the underlying Markov process could be modeled using a transition diagram to capture the changes in states. 
The formulation for the proof followed from the definitions of coefficient matrices that underpinned the transitions in state spaces ($\Lambda_{0}, \Lambda_{n}$). The constituent sub-matrices were then partitioned ($\Lambda_{0i}, \Lambda_{ni}, i=0,1...C$) to characterize for this coefficient matrix $\Lambda$ in equation \eqref{eq:8} and the equilibrium conditions $\Lambda P = 0$. Here $P$ denoted a column vector that defined for the probabilities that a queue was in a given state. \newrevise{Then this coefficient matrix formed the basis for derivation of expressions for the equilibrium conditions ($\Lambda P = 0$) of the queue sizes and}{This matrix $\Lambda$ was composed of sub-matrices (like $\Lambda_{0i}, \Lambda_{ni}, i = 0,1,\ldots C$) partitioned along the number of queues and there respective sizes. \newrevise{ the equivalent coefficient matrix to the equilibrium equations that defined the transition state space.. Eq.~\eqref{eq:8} defined for such a coefficient matrix ($\Lambda$) constituent of sub-matrices in the case of $C=2$ (under the assertion that the number of servers did not affect the formation of the coefficient matrix $\Lambda$)}{According to the authors, this matrix} was constituent of the regularity property and it was from this property that the solution emanated.  } It was therefore argued that the solution depended on choosing the proper sub-divisions. \revise{}{
\begin{equation} \label{eq:8}
\Lambda = \begin{bmatrix}
\lambda_{01} & \lambda_{02} & 0 & 0 & 0 & 0 & ....\\
0 & \lambda_{11} & \lambda_{12} & 0 & 0 & 0 & ....\\
0 & 0 & \lambda_{21} & \lambda_{22} & 0 & 0 & ....\\
0 & 0 &  0 & \lambda_{31} & \lambda_{32} & 0 &.....\\
\vdots & \vdots &\vdots &\vdots &\vdots &\vdots & \ddots\\
\end{bmatrix}
\end{equation}}
The matrices of varying dimensions were defined and elements partitioned into vectors to characterize specific queue state probabilities. This iterative technique was followed by an inversion of each matrix {\itshape ($\Lambda_{n2}$ and $\Lambda_{02}$)} to yield derivations for the closed form solution of all state probabilities. 
It was shown how the preference for a certain buffer when both were equal in size then became be a factor of the distance ($n_{i}-n_{k}\ge2$; {\itshape $i,k$ were queues}) the jockey candidate had to traverse to the alternative queue. It was also shown how a closed form solution could exist in scenarios where the coefficient matrix was dependent on the number of customers in a queue.

\newrevise{Findings from \cite{classica} inspired further exploratory work where \cite{Gen.Solutions} not only sought to formulate a general solution for the stability conditions of certain queue descriptors but also modeling}{The idea of \cite{classica} about sub-dividing the coefficient matrix motivated further exploratory work with queue descriptors as documented in \cite{Gen.Solutions}'s findings. The authors studied }how queue parameters like time-until-service and how often newcomers arrived in the system influenced the jockeying behavior. To switch to any shorter line was preconditioned on the size of one of the queues in an $M/M/C$ setup differing by two\newrevise{and a}{. Also, a}  different technique for the partitioning of the state space transitions into sub-matrices was utilized. 
The first model considered the case when no jockeying was permitted\newrevise{, expressions that characterized }{. The authors formulated }for the stability conditions and transition probabilities from the representational differential equations. Analytically, these differential equations were generalized as $AP=0$. $A$ denoted an $n(C^{2}-1)+1$ sized matrix whose elements corresponded to the weights (coefficient) of the state probabilities (where $n$ was the queue's capacity). Vector $P$ on the other hand denoted the actual probabilities that a queue was in a certain state \newrevise{ and}{ and} $0$ was a vector that constituted non-zero elements. It was this matrix $A$ that was sub-divided into a series of other column vectors and sub-matrices of varying dimensions \newrevise{ and}{. Then }by applying boundary constraints to the generalized equation $AP$, expressions that evaluated for the probabilities of the queue being a given state were formulated.
The second model then analyzed the jockeying behavior \newrevise{where the}{under recalibration in} queue descriptors like the arrival $\lambda$, the service $\mu$ rates and system utilization $\rho$ \newrevise{were a factor of}{relative to} the state of the a given queue. \newrevise{ hence}{This was  evident in} the variation in the representative matrix $A$ and magnitudes of the state probabilities. 
The performance results revealed that the time a customer had to wait until being processed was greater when jockeying was not allowed compared to when the behavior was permitted\newrevise{ reaffirming \cite{classica}'s findings}{. This was affirmative of \cite{classica}'s findings about the benefits associated to such impatient behavior}. \newrevise{Also, there }{Additionally, it was observed that there} was a higher probability of the queues staying idle under the non-jockeying setup. Other performance measures like the predicted system occupancy under varying degrees of the system utilization were documented. 

\newrevise{\cite{Zhao1990TheSQ} caused case for contention about the approaches used by researchers in most of the above reviewed findings}{\cite{Zhao1990TheSQ} was contentious about the techniques in the preceding studies}, arguing that besides the approaches being repetitive, \newrevise{they yielded no equations that evaluated for the boundary probabilities}{the solution ignored boundary conditions when defining the state space transitions}.
In their $M/M/C$ model therefore, the findings were aimed at unraveling some hidden dependencies between the transition rate matrix $R$ and the overall queue utilization. New admissions to the shorter of $C \ge 2$ queues tended to a Poisson distribution with rate $\lambda$. Each server processed jobs at \ac{IID} rates $\mu_{1}, \mu_{2}, \ldots,\mu_{n}$. \newrevise{}{In infinite time, the changes ($S=\{\overrightarrow{i}=(i_{1},\ldots,i_{C}|i_{i}\ge 0)$ for $1\le j\le C\}$ and $|i_{k}-i_{l}|\le 1$) in the server sizes} were generalized as a stochastic process ($\{(X_{1}(t)), X_{2}(t)), \ldots,X_{C}(t)), t\ge 0\}$). Here $k$ and $l$ denoted the number of customers in given states and stable conditions for this process were defined as $p_{\overrightarrow{i}}=\lim_{n \to \infty}P_{\overrightarrow{i}}(t), \overrightarrow{i} \in S$.
The generator matrix $Q$ and its partitions (sub-matrices $A_{ij}$ with varying dimensions) resulted from arranging the state space, that is, the ordering of state transitions $i$ and $j$. This was based on a function that defined which state came prior to or after another. 
The distributions of $\overrightarrow{p}_{i}$  were said to each constitute $2^{n}-1$ elements as states that formed a block $i$ with $P_{0}, \overrightarrow{p}_{1}$ as the bounds. \newrevise{The proof for the solution adopted mathematical techniques (separation of variables plus differential equations calculus)\cite{Mickens}\cite{Morse}}{From this evolved the difference equations essential} for the derivation of the equilibrium probabilities as a factor of the traffic intensity $\rho$. Taking the case of $i\ge2$ (therefore only interested in the $2^{n}-2$ probabilities in $\overrightarrow{p}_{2}$ since the bounds were known), it was shown that a solution only existed under steady conditions defined by $\overrightarrow{p}_{i+1}=\overrightarrow{p}_{2}\omega^{i-1}, i\ge1$ only when {\itshape $\det(A_{0}+A_{1}\omega+A_{2}\omega^{2})=0,$ $0<|\omega|<1$ (where $\omega=\rho^{C}$)}. \newrevise{And it was proven through a couple of theorems (which theorized that }{ Then building on the theoretical properties of } the eigenvalue(s) $r_{2}^{C}-1$ of the rate matrix {\itshape R} in the determinant, it was shown that $\rho^{C}$ was the only quantity that held true under this conditionality. These theorems were the basis for evaluating for the boundary and equilibrium probabilities of $p(i_{1},...,i_{C})$, $\overrightarrow{p}_{i+1}$(for block {\itshape i+1} ) and $\overrightarrow{p}_{2}$ to yield a relation between the rate matrix and the equilibrium probabilities.
Then taking the case of $C$ servers, to show that $p^{C}$ was the lone zero in the determinant above, a new chain for the stochastic process ($\Bar{S}=\{(i_{1},i_{2},....,i_{C})|i_{t}=0$ or $1, 1\le t\le C\}\bigcup \{m|m>C\}$) consisting of $2^{C}$ states was defined. \newrevise{and t}{T}he associated transition state infinitesimal generator matrix $\Bar{Q}$ was also constructed. Most ($2^{C}-1$) of the columns in both generator matrices ($Q $ and $\Bar{Q}$) were similar except for some few states (for which a specific generator matrix was formulated). The proof showed that for any server $k$ in the server deck $n$, $(\frac{p^{n}}{\omega})_{i}=constant,$ $ i\ge1$ if and only if $p^{C}=\omega$ ($\omega$ as the only real solution of the determinant). 
\newrevise{Expressions that characterized for measures in}{It was from these derivations that quantitative measures in} the average system occupancy, the time jobs \newrevise{spent being serviced to completion followed from these derivations and the application of Little's Law. The paper then}{took until exiting the system, were characterized for.} \newrevise{ shared}{The formulations were validated using} numerical experiments under varying configurations of the system descriptors.

Unlike single-customer impatience models, \cite{Martin2008AnalysisOJ} studied bulk workload migration. Arriving jobs were tagged with an expiry time and, if not processed before this time lapsed, were transferred to other servers. The transfer was defined by a policy chosen to minimize holding and migration costs. Arrivals obeyed a Poisson distribution while the service times were exponentially distributed. The paper first analyzed the case of a single-server fallback (where expired jobs reneged) and then generalized to $N$ servers with a transfer probability matrix $Q=(q_{i,j})$. Transfer rates take the form $\beta_i q_{i,j}$ and stationary probabilities $p_{i,j}$ (where $i$ was the server status of up or down and $j$ as the number of jobs) are derived for key cases. An iterative Poisson approximation was used to compute transfer rates and total cost, but the search for optimal $Q$ becomes combinatorial hard as $N$ grew. The authors therefore evaluated heuristic rules (round-robin, fastest-other, etc.) and reported numerical comparisons showing when heuristics approached optimal performance.

More like an extension to his \cite{2} earlier work, \cite{A.M.K.} investigated the evolution in the transition state space of an asymmetric $M/M/2$ setup of finite capacity $L$\newrevise{ using}{. Therein,} two theoretic propositions (randomisation theorem and Runge-Kutta) were essential \newrevise{statistical equations that characterized}{in the  formulation for} the  the dynamics in the state space. Jockeying was only allowed from the longer line only when any of the other service line was empty. The probability of having a certain number of customers at a time $t$ ($p_{i,j}=Pr(N_{1}(t)=i,N_{2}(t)=j)$) in either service lines was defined in a series of difference equations (\cite{A.M.K.}, Equations 1 - 21). 
The formulation of the model stemmed from the definition of the state probability vectors $P_{k}(t)$ (and their derivatives $P^{'}_{k}(t)$, $k=0,1,2,3, \ldots, L$) for all state transitions over $L$ during this time $t$. The difference equations were then re-defined in terms of these state probability vectors\newrevise{ to evolve into a}{. These equations were re-arranged into a} generalized block-matrix formation $P^{'}(t)=QP(t)$ \textit{(where $P(t)=(P_{0}(t),P_{1}(t), P_{2}(t),\ldots, P_{L}(t))^{T}$)} constituted by sub-matrices $(A, B, C)$. \newrevise{The}{The} sub-matrices denoted state space partitions to compose the generator matrix $Q$ as:
\begin{equation}
Q = \begin{bmatrix}
B_{0} & C_{1} & 0 & ... &  & 0 \\
A_{0} & B_{1} & C_{2} & 0 & 0 & 0\\
    & A_{1} & B_{2} & C_{3} & 0  & 0 \\
    &   &   &   & ... & \\
0 & .. &  & 0 & A_{L-1} & B_{L}\\
\end{bmatrix}
\end{equation}
It was argued therein that, the expressions for the rate matrices ($R_{L}=B_{L}^{-1}$, $R_{k}$) could be ascertained from iterations of computations\newrevise{, a method that deviated}{. This approach was in deviation} from the usual technique of calculating for the eigenvalues and eigenvectors of the rate matrix. The modified vector-geometric solution for the equilibrium probabilities of the stochastic process emerged therefore from the evaluations of the block-matrix equations in relation \newrevise{to the author's method of expressing for }{}the rate matrix given $P^{'}(t)={0}$.
\newrevise{The resolution for the}{And the solution for state} transition probabilities ($P_{i,j}(t)$) in finite state space followed from the randomization theorem \cite[Theorem 3.1]{A.M.K.}\newrevise{, which was more statistically efficient when computing for probabilities in state changes.}{.}
\newrevise{From applying this theorem to the difference equations emanated recurrence expressions whose properties were used to calculate for the distribution of state changes.}{From the subsequent application of the same theorem to the difference equations emanated the recurrence expressions. And based on the properties of these recurring expressions, the distribution of the state transitions was formulated for.}
It was also shown how the Runge-Kutta method could be manipulated to provide an equivalent evaluation for the transition state distribution. The work then provided numerical and comparative analyses \newrevise{(with \cite{Conolly.B}) of the results from both methods when used for the calculation of the probabilities in state changes, probabilities that the queue was empty, distributions about the entire system occupancy etc}{of both methods used in formulating for the densities of selected descriptors like the system occupancy}. \newrevise{The author was also interested in tests that sought to understand}{The studies also assessed} how the probabilities and capacities of the queues were influenced by the overall system utilization $\rho$ under variations in queue parameters. Also, the impact of switching queues on the average processing times and sizes of the queues was analyzed versus when no switching queues was possible.

\cite{11} studied the jockeying behavior and it's applicability to packet routing in multi-beam satellite systems. \newrevise{}{The} earth-stations were ordered as disjoint zones to form up-link and down-link connections and the sequence of the incoming packets was defined by an independent distribution function\newrevise{ such that they}{. New packets} were appended to the end of any of the shortest buffers at the satellites\newrevise{ and }{. They were} then processed at varying Markovian service rates following the \ac{FCFS} rule.
The vector-geometric solution was based on the assumption that the underlying process that characterized for such behavior was non-Markovian. Therefore, the process was segmented to first define an "imbedded Markov chain" ($\{\overrightarrow{X}_{l}=(X_{1}(t_{l}), X_{2}(t_{l}),....,X_{c}(t_{l})) l=1,2,....\}$)\newrevise{ for which}{. Then} expressions for the probabilities of the buffers' capacities for this chain were ascertained under ergodic conditions. The solution therefore  for stable conditions of the buffers arose from dividing the state spaces into groups $B_{<r}, B_{m}$ (maximum size of the buffers and maximum deviation between smallest and largest buffer respectively)\newrevise{ given}{. The division was premised on the } the probability ($\omega$) that there existed only a one-to-one relationship between any two states (with the exception of the boundary states) and that new packets did not necessarily visit all states. Following the sub-division of the state transition matrix was the formulation of the equilibrium equations (both for non-boundary and the boundaries states) that characterized for these sub-divisions. The solution for the queue equations borrowed from earlier findings (\cite{neuts1994matrix}, lemmas 1.2.4)\newrevise{that}{. That is, } there existed an eigenvalue $\omega = \omega_{0}; $  $ 0<\omega_{0}<1$ of the transition rate matrix $R$ and its determinant $\mathrm{det}(\omega I-\sum^{\infty}_{k=0}\omega^{k}A_{k})$ was 0. \newrevise{\textit{where denoted an Identity matrix and $A_{k}$ was a square sub-matrix block of states when there were $k$ packets in the system.}}{ Here $I$ denoted an Identity matrix and $A_{k}$ was a square sub-matrix block of states when there were $k$ packets in the system.} 
\newrevise{This same Lemma was adopted for the proof}{This lemma was fundamental to ascertaining} whether the vector-geometric solution was valid for the boundary equations \newrevise{ by taking}{. This was by simply taking} the probabilities of any of the boundary states ($p_{r,r,...,r}$) in $\overrightarrow{p}_{r}$ and evaluating \newrevise{for redundancy of each of the resultant substitutions of the vector-geometric expressions in the }{each for redundancy in the} equilibrium equations. \newrevise{The proof concluded by showing that, for}{ For} the stationary probabilities of this "imbedded Markov chain" $X_{1}$ therefore, there existed a geometric parameter $\omega =\sigma $ that defined the uniqueness of the solution. \newrevise{It was also proven that,}{Relatively,} the stationary probabilities vector ($\pi_{\overrightarrow{i}}=\lim_{t \to \infty}P\{\overrightarrow{X}(t)=\overrightarrow{i}\}, \overrightarrow{i}\in S$, $S$ denoted the state space) of the buffer capacities was also a modified vector-geometric solution governed by similar uniqueness constraints.
\newrevise{This followed by the definition of }{The studies furthermore defined} another "imbedded semi-Markov chain" ($X^{*}_{i}(t),$ $i=1,2,\ldots,c$) that represented the buffer capacities prior to any last packet at any time. \newrevise{Then, the fact that}{It was discovered that} the stationary probabilities for this kind of process were similar to those of the "imbedded markov chain"\newrevise{, it was theorized that from }{. Then, from } the sub-division of the stationary probability vectors $\overrightarrow{\pi}=(\overrightarrow{\pi}_{<r}, \overrightarrow{\pi}_{r},\overrightarrow{\pi}_{r+1},\ldots \ldots)$, \newrevise{}{it was shown that } the process assumed a modified vector-geometric solution too. \newrevise{Numerical experiments were conducted with the purpose of evaluating the system. The}{Numerical evaluations of the } performance when jockeying was permitted versus when this behavior was prohibited were documented\newrevise{ and }{. The} performance results showed how the time the packets spent in the buffers varied relative to the processing times, justifying the positive effects of this impatience behavior. 

\newrevise{}{The application of impatience in resource allocation was amplified by \cite{6}.} \newrevise{\cite{6}}{The work} derived for optimal rule-sets that controlled packets in an $M/M/2$ setup of \newrevise{ infinite capacity buffers of multi-beam satellite stations for cost effectiveness}{multi-beam satellite stations as buffer with infinite capacity}. Every packet that joined the server incurred holding costs and moving (instantaneous) a packet from one lane to another generated a jockeying cost but no preemption was allowed. 
Controlling \newrevise{of packets behavior (admission or jockeying)}{the admission or jockeying behavior} was also conditioned on the state of current service station\newrevise{ in terms of how much}{. This state was defined by the magnitude of the} load and the monotonic properties of the function $F_{ij}$\newrevise{(which defined the optimal rule-set as a factor of the expected accumulation of costs under discounted or non-reduced service costs and long-run average costs)}{. This control function $F_{ij}$(optimal rule-set) encapsulated the accumulated costs under discounted or non-discounted service costs and long-run average costs}. Therefore, a packet would only be routed to or migrated to another station if that station was in a valid state ($x_{1},x_{2}$) at time $t$.
In the model, \newrevise{the system changed states ($X(t)=(X_{1}(t),X_{2}(t))$ (state for lanes 1 and 2 respectively) then actions were taken when a new packet arrived or left any of the lanes}{ actions were state dependent; that is, given an increase or decrease in the size of either queues}.  
\newrevise{The proof proposed theorems for}{Basically, the work argued for} the existence of a function $F(x)$ that defined when it was okay for a new packet to be routed to a given queue, \newrevise{and when not to}{and when or not to move packets.} \newrevise{And the}{These} functions were then adopted in the characterization of the optimal rule-sets for the different behavior. 
Explicitly, a new packet was routed to lane 2 only if $x_{2}\le F(x_{1})$ was true. That is, if $F_{12}$ and $F_{21}$ were true then it was okay to move a packet from lane 1 to 2 ($x_{2}\le F_{12}(x_{1})$) or from lane 2 to lane 1 ($x_{2}\ge F_{21}(x_{1})$). Migrations of packets was not optimal when $  F_{12}(x_{1})<x_{1}<F_{21}(x_{1})$ and when $  F_{12}(x_{1})<x_{2}<F_{21}(x_{1})$.
One restrictive characteristic of the control functions was that the decision to move from a station with lower costs than its alternative was only plausible if the alternative station was idle. This characteristic was evaluated by taking a state of the system when the alternative station was not empty and validating it under the \newrevise{jockeying control that underpinned the best-fit rule-set}{optimal policy when jockeying was permitted}. 
\newrevise{The proof for existence of other}{The studies additionally formulated for the} asymptotic profiles of the control functions that defined the optimal rule-set $F$ under discounted cost. \newrevise{}{These asymptotic characteristics were were observed to be generic to the non-discounted mean costing over time.} Expressions for the expected reduction in costs $(V_{t}(x_{1},x_{2})$ over time $t$ and a predicted mean costs over longer time-span under steady conditions were also documented. It was ascertained though that, the control functions under both costing environments only converged under specific conditions of the both the jockeying and service costs.

\newrevise{Inspired by the notion of vendors offering varying pricing for their services to give the user more preference, according to \cite{wolfgangStadje},}{This notion of cost-oriented modeling to guide the impatience in queues was further studied by \cite{wolfgangStadje}.} \newrevise{it}{In their findings, it} did not matter whether the shorter or longer queue was joined or jockeyed to as long as the choice for either yielded costs below a preset limit $c$. These costs were a factor of both the size of the queue ($N_{i}$) and processing fee $\beta_{i}>0, (i\in\{1,2\}$, $i$ denoted a queue)\newrevise{, such that}{. A control function was also embedded to manage the admission of} new customers \newrevise{(with arrival rate $\lambda$ as a Poisson distribution)}{whose arrival periodicity obeyed a Poisson distribution with rate $\lambda$.} \newrevise{}{The arrivals } were routed to a queue therefore \newrevise{depending on the magnitude}{based on the measure} $\alpha_{i}N_{i}+\beta_{i}\le c$ (where $\alpha_{i}=1$ was a weighting measure on queue) of either queue. \newrevise{And staying away completely from the services when the overall service costs would turn up being too high was equally an option}{Alternatively, consumers had the option to stay away from the services when the overall service costs would turn out to be excessively high}. \newrevise{In a setup with finite capacity buffers of maximum length}{Each buffer eas finite in capacity to a maximum} defined by $K_{i}=[c-\beta_{i}+1]$\newrevise{, }{. The} representational Markov chain $X(t)=(X_{1}(t),X_{2}(t))$ of the changing queue sizes was irreducible and a factor of the underlying costs. 
Formulations for the steady-state distribution deriving from theoretical comparisons between selected queue descriptors ($K_{i}, \beta_{i}, c$) were defined under specific quantities of the system utilization ($\rho=\frac{\lambda}{2\mu}, \rho \neq 1, \rho \neq \frac{1}{2}$)\newrevise{ and}{. } \newrevise{for each comparison, expressions that represented the state changes (as a factor of $\beta_{1}$, $\beta_{2}$ and $c$) obtained. The distinction line of the three cases compared was based on the magnitude of $K_{1}-K_{2}$ (to determined whether a new client had to join or jockeying to either $server1$ or $server2$);}{The comparisons were performed for three cases that were distinguished based on the magnitude of $K_{1}-K_{2}$. This measure determined whether a new client had to join or jockeying to either $server1$ or $server2$.} \newrevise{That is,}{For each of the cases} $\beta_{2}-\beta_{1}-1<K_{1}-K_{2}<\beta_{2}-\beta_{1}$, $\beta_{2}-\beta_{1}<K_{1}-K_{2}<\beta_{2}-\beta_{1}+1$ and $K_{1}-K_{2}=\beta_{2}-\beta_{1}$ \newrevise{such that all possible states reachable for each of the three cases were defined}{ all possible states reachable were defined}. \newrevise{For each comparison case,}{And for each of these inequalities,} balance equations \newrevise{which characterized for the}{definitive of the} influence of leaving or entering the queue in a given state were shown. \newrevise{ and under}{Then taking into account} what circumstances a specific state was reachable \newrevise{formed the basis }{was essential} for the proof for equilibrium probabilities ($\pi_{i,j}$) of the system. The solutions for these balance equations evolved from them being re-written as difference equations in relation to the state sequence $s_{i}, i \in \mathbb{Z}_{+}$. It was shown from induction principles how $\pi(i,j)$ could be derived for from relations between $\pi(1,0)$ and $\pi(0,0)$.

Similar admission-control schemes route arrivals between two service lines (M/M/C, $C=2$) using Bernoulli-assigned probabilities and charge customers for waiting and for any jockeying between queues \cite{9729148}, building on earlier work \cite{3}. Customers continuously receive queue-state updates and decide whether to stay or switch so as to minimize expected cumulative cost; the system state is $x=(q_1,q_2,\ell)$ and actions $a\in\{0,1\}$ incur position-dependent costs $V_n,V'_n$ that evolve with time. The authors derive threshold-type (limit) policies: there exist queue-length bounds $q^*(q_1,\ell)$, $\ell^*(q_1,q_2)$ and an admission threshold $\varepsilon(q_1)$ such that a customer switches only when the alternative queue lies below (or above) these limits, and a new arrival prefers one queue when the other exceeds $\varepsilon(q_1)$. Monotonicity properties of the optimal rule set are proved to show that the limits are nondecreasing or nonincreasing in the relevant arguments. So both jockeying and admission rules reduce to simple threshold decisions in the long run — results that echo and refine the limit-policy analysis in \cite{DOWN2006509}.

\newrevise{For jockeying control, \cite{ravid_2021}'s work concentrated on obtaining}{More theoretical models to optimize  impatience among jockeying consumers was the subject of \cite{ravid_2021}'s work.} \newrevise{analytic expressions that evaluated}{The studies analytically expressed } for the frequency of jockeys made from one service station to another before getting served. New arrivals (obeyed a Poisson distribution) were pushed to the shorter station or one of the two stations with equal probability\newrevise{and the jockeying for a tail-end customer from the}{. Each jockeying activity accumulated a cost for moving from the tail-end of a}  more occupied station to the end of the less occupied station\newrevise{ was permitted when }{. This was given that }the difference between the length of any of $k \ge 2$ servers hit the preset threshold $d$. \newrevise{(each serving customers at \ac{IID} processing times)}{(Each station served waiting customers at \ac{IID} processing times.}
The \newrevise{solution involved splitting the}{} transition state space \newrevise{}{partitioning here was different.} \newrevise{}{It was }based on queue length statistics like customers ahead ($f$) or behind ($b$) a given customer ($(f,b)\implies b\ge f\ge 1 $) in the service line\newrevise{and }{. An additional constraint on the behavior was }whether the next move by that customer was a jockey or a forward in the same queue. 
\newrevise{The proof was adopted to provide }{Inspired by concepts } from generating function theory, \newrevise{sought}{ the objective function encapsulated}  mappings for a customer's position in the queue to the possible number of jockeys ($Y_{f,b}$) that the customer would make before being processed. The evaluation of the generating function \newrevise{therefore was characterized by the}{ was a procedural and } iterative formulation of the relationships ($\Phi_{f,b}(s)$ $0\le s\leq 0$) between these states and actions. This required initial statistical computation for the chance ($P_{f}(j|b),1 \le f \le \min{b,j}$) that the customer's next state ($f-1,f$) followed a jockey to the alternative queue or a forward move in the same queue; \newrevise{ to end up}{ Such that the customer ended up} in state $(f-1,f) f \le \min {j,b}$. 
The expression for the chance that a customer would move to the alternative station (in state $(f-1,j), f \le j \le b-1$ or $j \le b$) were premised on the prevailing conditions in the current queue. That is, provided that $b>f-1$ and that the variation between the number of clients behind the current customer that left that station versus the number that wanted to join any of the queues was $b-j$. 
\newrevise{From the total probability \cite[Equation 1]{ravid_2021} principles, the formulation of the relations (given the probabilities of being in a state) evolved into Eq.~\eqref{eq:44} which defined for the generator function ($\Phi_{Y}(S)$).}{Equation \eqref{eq:44} was this generator function that characterized for the conditions, state probabilities and relations.} This function mapped the expected number of jockeys ($Y_{f,b}$) from one queue to another before the customer was serviced\newrevise{ and it was a prediction }{. This frequency was }dependent on $f$ (the number of people ahead of the current customer being served) and the arrival rate. 
\begin{equation}
    \Phi_{Y}(S)= \frac{2-\rho}{2+\rho}\Bigg[\frac{2+\rho}{2}+2B_{0}(S)+\frac{4}{\rho}B_{1}(S)\Bigg] \\
    \label{eq:44}
\end{equation}
\textit{where $B_{0}$ and $B_{1}$ were linear expressions that denoted aggregations of state relations and service line utilization over n customers in a queue. And $0\leq s \leq 1$}. 

\noindent Additionally, a generating function ($\Psi_{K}(\theta)$) that yielded the random distribution of how many customers $K$ ahead of the current customer left the queue was defined. This followed the analysis of the process representative of either actions (joining or leaving) over time, therein referred to as a \textit{"difference random walk" (DRW)}.

The authors in \cite{9729148} combined static and dynamic admission, service and jockeying controls to derive an equilibrium “hedge-point” policy. Here, incoming jobs did not pre-select a queue but were routed by an admission controller \cite{3}. Each queue operated at exponential service times, and the admission rule $u\in\mathcal U$ mapped system state $x$ to actions that traded off holding, service and jockeying costs via a discounted value $J^u(x)$. For optimal control, the value function $ V(x) = \max\limits_{u \in U}J_{u}(x)$ had to be maximized. This was by iterating a Bellman operator $T$ and retaining only value functions with the required structural properties (sub-modularity or concavity) so that the operator converged. The resulting optimal policy was threshold-style and described by switching functions $S_1,S_2,L_i,G_i$ that governed admissions, routing and migrations. The numerical value-iteration experiments for symmetric servers (equal costs and rates) illustrated the approach and the convergent approximation of $V(x)$.

\newrevise{\cite{Jiang.Tao.Liu} is built on earlier findings}{Building on earlier findings} \cite{Haque,Miyazawa}, \newrevise{but with specific interest in the boundary asymptotic formation of the probability distributions of the queue sizes in an $M/M/2$ system (one special buffer and the other normal)}{\cite{Jiang.Tao.Liu}'s work focused on the asymptotic properties of the boundary conditions of the queue length distributions. In an $M/M/2$ setup, the aim was to formulate for the tail probabilities of the two heterogeneous queues}. The growth in size of the main buffer was continuously monitored at exponential time intervals\newrevise{such that workload was transferred ($L-K$ where \textit{L} was the length of the normal line) when the size of the line exceeded a preset threshold $K)$}{. And when the size $L$ of this main queue exceeded a preset threshold $K)$, workload was transferred to the auxiliary queue}.  
For the representative continuous Markov chain $\{(L_{1}(t), L_{2}(t)), t\ge0\}$, steady-state conditions of the queue sizes were first derived for as $\lambda q<\mu_{2}$ and $\lambda<\mu_{1}+\mu_{2}$\newrevise{\textit{(where $\lambda<\mu_{1}, \lambda<\mu_{2}$ were arrival rates and $q$ the probability that new arrivals joined the special line)}}{. Here, $\lambda<\mu_{1}, \lambda<\mu_{2}$ denoted the arrival rates and $q$ the probability that new arrivals joined the auxiliary buffer)}. The proof developed from adoption of the Foster-Lyapunov\cite{Foster}\cite{Foster-Lyapunov} condition for stability to show that the some process states were revisited in finite time (positive recurrence). 
For uniformity, the transition probability matrix $P=I+Q$ (\textit{where Q represented that rates at which the queue varied in size and $I$ an Identity matrix}) was subsequently divided into sub-matrices based on levels. \newrevise{By exploiting the special structural properties of the irreducible sub-matrix ($D(\sigma)=A+B\sigma+C\sigma^{2}$ - \textit{where $\sigma$ defined by Eq.~\eqref{eq:55} was the decay rate; A,B and C as sub-matrices depicting state space splits}), the authors showed that the boundary limits of the probability distribution of the normal queue size decreased at some constant ratio sequentially (geometrically).}{It was observed that one of the sub-matrices $D(\sigma)=A+B\sigma+C\sigma^{2}$ (\textit{where $\sigma$ as defined by Eq.~\eqref{eq:55} was the decay rate, and A,B and C were sub-matrices depicting state space splits}) was irreducible and inherent of special spectral properties. \newrevise{from which it was conclusively generalized that the tail characteristics of the distribution of the main queue size decreased at some constant ratio sequentially (geometrically). A further split of the sub-matrix $D$ along the boundaries evolved into the formulation for the proof. }{These properties were fundamental to the derivations, such that a further split of the sub-matrix $D$ along the boundaries was followed with evaluations for its convergence norms $\overrightarrow{\sigma}, 0<\sigma<1$; And verification for the existence of the $\frac{1}{\sigma}-invariant$ measures \cite[Lemmas 4.1-4.2]{Jiang.Tao.Liu}.}
\begin{equation}
    \sigma=\frac{(\lambda p+\mu_{1}+\eta)-\sqrt{(\lambda p+\mu_{1}+\eta)^2-4\lambda p \mu_{1}}}{2\mu_{1}}
    \label{eq:55}
\end{equation}
\textit{where $p$ was the probability of joining the main queue, $\eta$ was a distribution of the server polling intervals}.

\noindent Specifically, when $\lambda q<\mu_{2}$ and $\lambda<\mu_{1}+\mu_{2}$, it was shown that the asymptotic properties for joint stationary distribution $\pi_{i,j}$ ($i,j$ denoted the level and phase in the state space composition) as approximated from the decay rate decreased geometrically. \newrevise{That is,}{In essence, it was conclusive that, the decay rate varied with the individual queue sizes such that } the number of customers in the main line decreased at some constant ratio in infinite time\newrevise{ vis-a-vis}{. This was contrary to} the number of customers in the auxiliary line, which stayed constant.} 
Numerical analyses of the these formulations were performed, experimenting with different queue parameter settings for the arrival or processing rates in each queue. At different monitoring intervals, results were compared under changing values of the arrival rate and it was observable that the increase in the decaying rate was linearly proportionally to the rate at which customers joined the queue. It was interesting to also observe how the decaying rate responded to increasing quantities of the processing rates\newrevise{ and the results surprisingly suggested that}{. Of performance significance was the observation that} the decaying rate decreased linearly with respect to an increase in the service rates of the queues.

\newrevise{}{\cite{Rosberg_etal}'s work proposed an alternative energy-efficient heuristic for guiding efficient use of energy when allocating jobs in multi-server processor sharing setups. The setup was composed of heterogeneous infrastructure with finite buffer sizes. The job streams followed a Poisson distribution at rate $\lambda$ to land on any of $j \leq n$ servers for processing at exponentially distributed rates. Each server consumed energy ($\varepsilon$) at a rate ( defined by \eqref{eqn_ee}) that was monotonically decreasing with the service rate. The object of the studies was to develop energy efficiency policies when apportioning workload while maximizing throughput.
\begin{equation}
    \varepsilon (\mu) = \mu^{3}
    \label{eqn_ee}
\end{equation}
\textit{where $\mu$ denoted the node's processing speed.}

In the studies, two policies were benchmarked. The baseline heuristic was the insensitive jockeying policy, therein referred to as slowest-server-first (SSF). Jockeyed jobs displaced existing jobs backward or forward and position allocations for jockeys were defined with equal probability to departures. 
The proposed energy-efficient (EE) rule-set on the other hand allocated tasks to the first $\hat{n}\leq n$ set of busy servers such that the tasks were routed to the least occupied or empty buffers. Then the next $\frac{s}{b}$ servers were selected for task processing if all instances in the $\hat{n}$ set were occupied.  Here, $b$ defined for the finite size of a particular queue and the state space $s \in S$ denoted the number of jobs in the queue. The energy efficiency of the servers was then collectively calculated as the ratio of the summed long-run mean throughput $T$ and the expected consumed power $E$ as $\frac{T}{E}$. It was revealed that this server cascading (in states $\hat{n}<s\leq \hat{n}b,$ $s \in S$) yielded relatively lower task processing times to minimize the overall load and balance server utilization.
Formulations for comparative analysis of the two heuristics followed from the theoretical propositions that showed conditions for $\hat{n}, b, \eta$ and $\mu_{j}$ ($\hat{n}\geq 1$) under which the energy-efficient policy was more optimal than the SSF policy. Qualitative measures were then defined by the relative error between the two rule-sets computed using \eqref{eqn_error}. Numerical evaluations of this error were performed in selected quantities of $\eta$ and $\mu$ in a series of experiments. This was purposely to determine the optimal $\hat{n}$ that maximized the servers’ energy use ($\frac{T}{E}$) in the entire set.
\begin{equation}
    \Delta \hat{E}^{SSF}_{EE_{\hat{n}}} = \Delta \left( \frac{T_{EE_{\hat{n}}}}{E_{EE_{\hat{n}}}}, \frac{T_{SSF}}{E_{SSF}} \right )
    \label{eqn_error}
\end{equation}

\noindent It was shown that, a certain measure of $\hat{n}_{*}$ provided a ceiling such that the mean aggregated processing rates should never exceed the mean arrival rates. Under preset configurations of $b, a_{i}$ (where $a_{i}$ was a measure of how two cascaded server groups differed in processing capacity), the EE policy was more optimal energy-wise than the SSF. However, this efficiency decreased with server utilization or traffic intensity. This meant that EE was only more optimal under limited traffic conditions. Under similar settings though, no significant improvements were recorded in the throughput of the EE ruleset.
}

\paragraph{\revthree{\textbf{Limitations of Statistical Models in Next Generation Networks:} Statistical modeling of jockeying in queues faces several practical limits in next-generation networks. For example, in such environments telemetry is often partial, delayed or censored, producing biased and high-variance estimates. Also, the system is very dynamic (mobility, auto-scaling, flash crowds) causing rapid concept drifts. New architectural abstractions like slice or tenant heterogeneity invalidates pooled models. And rare but useful tail events like SLA breaches, are underrepresented by standard loss functions. closed-loop effects (predictions changing behaviour) and multi-agent interactions break offline validity; edge compute and latency constraints limit model complexity and freshness; uncertainty quantification is frequently absent or miscalibrated; telemetry can be adversarially manipulated; and ground-truth labels for “beneficial” jockeying are ambiguous—altogether making naïve statistical predictors fragile unless paired with censoring-aware estimators, adaptive retraining, causal or closed-loop evaluation, and robust telemetry authentication.
}
}

\subsection{\revise{}{Nash Equilibrium based Models}}
\newrevise{Nash-equilibrium rules for an $M/M/2$ system with threshold jockeying permitted were studied by \cite{Hassin.Haviv} to understand the value of prior purchased queue status information to a customer. The experiments were inspired by }{Nash equilibrium rules have also been applied in studies that model for impatient in buffering systems. Take the case of \cite{Hassin.Haviv} who was interested in how valuable prior queue status information was to the prospective consumer before joining the system.} \newrevise{}{The experiment in an $M/M/C=2$ setup were motivated by }the notion that such information underpinned optimal usage of the service lines to consequently minimize waiting times \newrevise{and that }{. This meant that,} for new arrivals, preference for which queue to join was influenced by the precedent customer having bought similar information (externalities). \newrevise{}{Basically, on arrival a customer purchased a probability value that abstracted the potential benefits given the current state of the queue.} The authors sought to ascertain whether this information from prior clients impacted the subsequent client positively or negatively.
Analogous to a cost benefit model, the authors deployed Nash-equilibrium strategies that put a value on the purchased status information by evaluating how much benefit a client got from it. Two strategies arose here $p, 0\le p\le 1$ which denoted the probability whether the information was purchased or not respectively. That is, a pure and a mixed strategy. But the strategies ($g(p) = C, g(p)\le C, g(p)\ge C$) were a factor of the relation between the benefit of the acquired information $g(p)$ and the costs $C$. The expected benefit $g(p)$ meant knowing how much less time the consumer would be waiting to get serviced given the charges for that information.
Following a partitioning of the state space, the authors used the matrix-geometric method \cite{neuts1994matrix} to obtain the stationary probabilities $\pi_{i,j}$ \textit{(i or j being sizes of either queues)} of each queue size given the jockeying threshold $N=3( 3\le N\le\infty$). The stationary probabilities \newrevise{were evaluated}{were then characterized for} from the eigenvalues/eigenvectors and spectral properties of the rate matrix $R$.  
For each assumed position in the therefore, a function $g(p)$ for computing the benefit (expected waiting time) at that position under a given Nash-Equilibrium strategy\newrevise{ was formulated from the difference equations}{}. 
The numerical results \newrevise{showed comparisons between}{compared} the benefit from purchases \newrevise{depending on }{under varying measures in }the jockeying threshold $N$ and the service line utilization $\rho=\frac{\lambda}{2\mu}$ (\textit{where $\lambda$, $\mu$ were the arrival and service rate respectively}). 
The study conclusively expressed for the magnitude of influence (be it negative or positive) of the actions of prior customers on new arriving customers when they purchased knowledge\newrevise{and when $N=3$}{}. The effect was considered positive if the acquired knowledge helped the consumers optimally use the service line and negative if the customer ended up waiting longer than expected. \newrevise{Evaluations }{Performance evaluations }for the average sojourn time under varying magnitudes in system occupancy revealed that as $N$ grew larger than 4, the benefits of purchasing knowledge were negligible. 

\paragraph{\revthree{\textbf{Limitations of Nash Equilibrium Models in Next Generation Networks:} Nash models assume well-specified utilities and substantial knowledge (or common priors) about other players, assumptions that fail under partial, delayed, or privacy-constrained telemetry in slice-based MEC or 6G settings. They also require fully rational agents and often slow iterative dynamics; in practice agents are bounded-rational, run learned heuristics, and environments change too quickly for equilibria to form. Scalability is problematic—computing equilibria for many tenants, slices and cross-domain constraints is intractable—and enforcement across administrative domains (migration budgets, billing, penalties) is nontrivial. Finally, standard NE ignores adversarial misreporting. Mitigations include Bayesian/mean-field formulations, bounded-rationality or learning dynamics, mechanism design for enforceability, and trace-driven validation, but Nash Equilibrium techniques are mainly useful as a baseline or for small, slow subsystems.}
}

\subsection{\revise{}{Fluid Theory based Models}}

\cite{DOWN2006509}'s fluid model analyses extend jockeying studies by treating bulk arrivals and transfers as continuous flows. In an $M/M/2$ setting with periodic Poisson batches, controllers decided bulk transfers between a low-cost and a high-cost service line to minimize long-run costs (holding plus fixed/variable transfer costs) while avoiding idling. 
Equation \eqref{eq:34} characterized for the average predicted cost of a client starting in a state $x$. Then adopting a given rule-set $\pi$ as $g^{\pi^{*}}(x)$, the optimal rule-set $\pi^{*}$ was conditioned on $g^{\pi^{*}}(x)\le g^{\pi}(x)$ for all states. 
\begin{equation}
     g^{\pi}(x)=\lim_{n\to\infty} sup \frac{\mathbb{E}^{\pi}_{x}\bigg\{\sum^{n}_{i=0}[ k(X_{i},Y_{i})+\int^{\rho_{i}+1}_{\rho_{i}} c(X_{i},Y_{i})\delta t] \bigg\}}{\mathbb{E}^{\pi}_{x}\{\rho_{n}\}}
   \label{eq:34}
 \end{equation}
\textit{where $X_{i}$ was the state at the $i^{th}$ decision step and $\mathbb{E}^{\pi}_{x}$ the expectation (given policy $\pi$) that taking action $Y_{i}$ would accumulate $k(.)$ as the overall cost at a cost rate $c(.)$} 
Policies $\pi$ specified the transfer magnitudes and timings and each policies were evaluated by their asymptotic average cost. The optimal rules were then those whose sample-path cost $g_\pi(x)$ was minimal for every state. The fluid model was said to be in equilibrium when all incoming clients had been processed ($\Bar{M}(t)=0,$ $t\ge t_{0}$) by the system (\textit{where $\Bar{M}(t)$ defined for changing system occupancy given instantaneous arrivals and service completions })\newrevise{ (service rate greater than the arrival rate) and}{. And} otherwise ($\Bar{M}(\delta)\neq 0;$ $\delta>0$) if at a certain time-frame $\delta>0$ the queues were still occupied. The theory yielded threshold/non-idling conditions under which bulk migrations were justified (move to the costly line only to prevent idling), and enforced symmetry constraints for single or bulk transfers. Numerical experiments for asymmetric multi-queue setups (varying costs and arrival rates) showed how polling interval $T$, transfer granularity and policy choice affected total cost and system load. And confirmed the benefits of carefully parameterized bulk-transfer rules. These results generalize earlier discrete-control findings by \cite{Koole} while highlighting computational/design tradeoffs.

In \cite{Delgado2014StabilityAO}'s studies, fluid theory was the basis for formulating steady-state expressions for a cluster of cascading servers\newrevise{that cooperated}{. The endpoints were stacked so that they collaborated} to share task executions. Specific classes of customers with \ac{IID} arrival times were assigned positions in specific queues. But these class specific queues served customers from other classes only when they were idle \newrevise{}{at \ac{IID} processing times}. 
Two variations in modeling were presented therein\newrevise{;}{. O}ne of the class of \textit{X-models} as a setup with two servers \newrevise{ (}{, such that}each represented a class of customers\newrevise{and the other}{. The other model on the contrary}, \newrevise{therein coined the term}{therein referred to as a}  "tree-cascade system", was defined with three servers (classes).
It was argued that to compute the steady-state conditions for such a networked system, one needed to prove the stability of the underlying fluid limit model\newrevise{, which required that the system was in equilibrium}{. necessary for these conditions was} the emergent Markov process $X$ having a non-repeating consistent statistical value \cite{10.2307/2245268}. \newrevise{It was theorized (from \cite{DELGADO201439})}{Theoretically therefore, as earlier suggested by \cite{DELGADO201439}}, given \newrevise{two service lines, where one had }{a service line with }a lower processing rate ($r_{1},r_{2}\ge 1$ or $r_{1},r_{2}<1$) than the other (given $r_{1}=\frac{\mu_{1}}{\mu_{2,1}}$ and $r_{2}=\frac{\mu_{2}}{\mu_{1,2}}$), then stability for fluid limit based models could only exist under specific conditions of comparative variations in the arrival ($\lambda$) and service rates ($\mu$). These conditions were specified by Eq.~\eqref{eq:51}, such that the aggregations over the  quantities $\Bar{Q}(0), \Bar{U}(0), \Bar{V}(0)$ of the queue network equations at a time $t\ge0$ equaled unit, $\Bar{Q}(t)=0,$ $t\ge t_{1}$.
\begin{equation}
  \begin{cases}
    (A_{1}) & \lambda_{1}-\mu_{1}+\frac{\lambda_{2}-\mu_{2}}{r_{2}}<0,\\
    (A_{2}) & \frac{\lambda_{1}-\mu_{1}}{r_{1}}+\lambda_{2}-\mu_{2}<0.
  \end{cases}
  \label{eq:51}
\end{equation}
\textit{where $U(t) =$ the time before a new arrival seeks to join the server, $V(t) =$ time left to service end for customers and $Q(t) = $ size of the buffer at any time.}

\noindent The proof for the stability of the $X-model$ then derived from the adoption of the Lyapunov function $f(t)$\cite{Foster-Lyapunov}. \newrevise{which related the sizes of the queues }{This function encapsulated the relationships between the sizes of the queues}($\Bar{Q}(1),\Bar{Q}(2)$) over time $t\ge0$\newrevise{and it was necessary to show that under varying comparisons of the arrival ($\lambda_{i}$) and service rates ($\mu_{1}$}{. Essentially, it was then necessary to show that under varying conditions of $\lambda_{i}$) and $\mu_{1},\mu_{2}$,( $1,2$ being queues), this function was bounded by the inequality} $f(t)\le -C$ ($C>0$ ). 
The studies were extended to the tree-cascade setup under work-conserving rules\newrevise{where }{. In this setup,} jockeying was allowed to any of the servers \newrevise{that were free although}{empty. However,} the third station was dedicated to supporting the other two stations.
The corresponding Markov process $X$\newrevise{(renewal arrivals and jockeying when one queue was empty)}{} was similar to the one for the two server setup \newrevise{except with higher dimensionality in the state space}{with the exception that the new state space had a higher dimensionality.} \newrevise{and with}{ The implication of this was }slight differences in the system equations that expressed for the interactions within the service stations.
The stability of the fluid model in this setup was also premised on ratios of processing times ($r_{1,3}\le r_{1,2}, r_{2,3}$) and the rate at which customers sought to join a given buffer. 
Similarly, \newrevise{adopting the Lyapunov function $f(t)$}{the respective Lyapunov function $f(t)$ was defined} \cite{Foster-Lyapunov}\newrevise{and because the}{. And this} function was inherent of comparable differentials at any time $t$ with the server sizes\newrevise{, the}{. The} proof for the equilibrium conditions then gathered from evaluation of the inequality $f(t)\le-C<0;$ $C:=\min\{C_{1}, C_{2}, C_{3}\}$ to hold under varying sizes of the queues ($\Bar{Q}_{i}(t), i=1,2,3$).
It was shown how the fluid limits on the servers' occupancy hold stable in infinite time\newrevise{, hence that with such server conditions the Markov process revisited specific states in finite time (positive Harris recurrent)}{. Hence, the conclusion that under such server conditions, the stochastic process was characterized by revisits to some specific states in measurable time, that is, positive Harris Recurrent}.

\paragraph{\revthree{\textbf{Limitations of Fluid Models in Next Generation Networks:} Fluid models average discrete events into continuous flows, which can obscure tail events and discrete triggers (spikes, deadlines, migration cascades) that drive SLA violations in MEC or 6G networks. These models also assume smooth, slowly varying inputs, yet abrupt network function scaling, fast mobility or rapid slice reconfigurations produce transitions not easy to encapsulate in fluid models. Hard constraints (migration budgets, per-task costs) and event-driven control logic are also awkward to represent, and fluid controllers can become unstable when feedback is delayed or partial. Fluid modeling however remains valuable for aggregate insight and baseline controller design but must be robustified for operational deployments.}
}

\revise{}{
\section{Analytic Models}
}
\subsection{\revise{}{Matrix-geometric models}}
Modeling of the shortest queue problem using matrix geometric approaches was pioneered by \cite{neuts1994matrix}. \newrevise{was foundational}{ This was the building block for} \cite{Gertsbakh1984TheSQ}'s experiments to expressed for the stationary probability vectors of queueing systems.
The study simulated an $M/M/C=2$ airport setup with planes that accessed runways (as queues) \newrevise{following}{at a frequency defined by} a Poisson distribution with rate $\lambda$\newrevise{ and }{. The time at which planes took off was} exponentially distributed with rate $\mu$ for each runway. 
\newrevise{The continuous-time Markov chain representation of the state transitions for all the dynamics within such a system took the form of an infinitesimal generator matrix \textit{Q} (Eq.~\eqref{eq:18}) with sub-matrices ($A_{0}, A_{1}, A_{2}$ and $B_{0}$) that inherently encoded state transitions}{The state space transitions evolving from all the dynamics within the systems were encoded by a continuous time markov chain. That encoding morphed into a form of an infinitesimal generator matrix \textit{Q} defined by Eq.\eqref{eq:18}. The matrix was fundamentally constituted of sub-matrices ($A_{0}, A_{1}, A_{2}$ and $B_{0}$) representative of the state transitions}. 
\begin{equation}
Q = 
\begin{bmatrix}
B_{0} & A_{0} & 0 & 0 & 0 & ....\\
A_{2} & A_{1} & A_{0} & 0 & ....\\
0 & A_{2} & A_{1} & A_{0} & 0 & ....\\
0 & 0 & A_{2} & A_{1} & A_{0} & 0 .....\\
.\\
\end{bmatrix}
 \label{eq:18}
\end{equation}
Ideally, it was suggested that completing a service in either queue and a jockey to a shorter queue summed to a total transition rate of  $2\mu$. \newrevise{And this also altered the}{This had the direct implication that, the} dimensions of the sub-matrices that composed the generator matrix \textit{Q}. \newrevise{}{It was therefore argued that, necessary steady state conditions of the stochastic process existed only if ${\lambda} < 2{\mu}$ and a redefinition for the a non-negative rate matrix $R$ that represented the rates at which states changed. This was premised on the proposition (\cite{Gertsbakh1984TheSQ}, proposition 1) that, given the underlying dynamics in such processes, there was a probability that certain queue states ($i,j$) in the process were revisited in finite time (positive recurrence \cite{neuts1994matrix}).}\newrevise{For that, a non-negative rate matrix $R$ that represented the rates at which states changed was defined and}{} 
\newrevise{It was then shown how the proof for the necessary steady state conditions of the stochastic process existed only if ${\lambda} < 2{\mu}$.}{}\newrevise{This was premised on the proposition (\cite{Gertsbakh1984TheSQ}, proposition 1) that the underlying process to such dynamics inherently possessed conditions or there was a probability that certain queue states ($i,j$) in the process were revisited in finite time (positive recurrence \cite{neuts1994matrix}) .}{}\newrevise{}{Subsequently expressions for the stationary probability vectors $\pi=(\pi_{1},.....\pi_{n})$ were formulated. }And the stationary probability vectors were verified to exist under the steady conditions ${\lambda} < 2{\mu}$. 
The work provided a numerical analysis comparing experiments with variations in parameter settings for \newrevise{how busy the queue was}{the queue intensity} $\rho$\newrevise{and $n(n=5, n=10, n =15 \ldots)$. }{. Under varying dimensions of the jockeying threshold $n(n=5, n=10, n =15 \ldots)$, it} was conclusively suggested that as the threshold $n$ increased $(10<n>\infty)$, differences in stationary probability vectors became negligible. The experiments also provided computations for some queue statistics like the means of individual queue lengths \newrevise{, mean waiting time}{and the mean waiting time.} \newrevise{and reaffirmed findings}{The findings were a further justification of earlier revelations \cite{11}} that systems where jockeying was permitted performed better than those where the behavior was prohibited.

\cite{KAO199067} extended the application of these matrix-geometric techniques to an \ac{M/M/C} queue setup. \newrevise{It was shown that from manipulating the structure of the generator matrix emanated a more reliable approach for ascertaining the stationary probability vectors. This followed the author's counter arguments about the methods used when partitioning the state space in \cite{Gen.Solutions}, hence suggestions for an alternative solution.}{This followed the author's counter arguments about the methods used when partitioning the state space in \cite{Gen.Solutions}. That, by manipulating the structure of the generator matrix, a more reliable approach for ascertaining the stationary probability vectors for the stochastic chain of events.} 
The \newrevise{setup}{experiments} assumed multiple servers with an infinite number of tenants that arrived following a Poisson distribution with rate $\lambda$\newrevise{and each}{. Each} server processed tenants at \newrevise{rates that were exponentially distributed}{an exponentially distributed rate $\mu$}. New arrivals joined the shortest queue but if the queues were equal in length, they joined either queues with the same probability\newrevise{ while jockeying}{. Jockeying on the other hand} was allowed only when the difference in queue sizes was two (jockeying threshold $n = 2$). The representation of such a \ac{QBD} process was an infinitesimal generator matrix $Q$ composed of sub-matrices that denote the state transitions of the queue lengths\newrevise{ but introduced slight modifications}{. However, slight modifications were introduced} to the structure of the stationary probability vectors.
\newrevise{It was proposed}{The justification of their argument followed from the proposition} that the boundary conditions required a special extension of the probability vectors, hence the need for re-expressing for the rate matrix $R$.
\newrevise{The proof applied previous theorems (theorem 3 \cite{GILLENT1983151}) such that the iterative evaluations culminated into an expression for the rate matrix \textit{R}}{And theoretical findings (theorem 3 \cite{GILLENT1983151}) were fundamental for the formulation for this rate matrix $R$}. 
\newrevise{Extensions were also done to \cite{Ramswami1986AGC}'s work, introducing the argument }{The studies also revisited  \cite{Ramswami1986AGC}'s work, arguing } that because the structure of the infinitesimal generator matrix exhibited the likelihood of boundaries state being revisited, there was need for new \newrevise{evaluations}{definitions} of the average size \textit{L} of the queue. \newrevise{}{Therefore, the representative infinitesimal generator matrix was first re-partitioned. }
Then, it was proposed that, as part of the state space of the underlying \ac{QBD} was an absorbing state $\theta$\newrevise{(a state when a new arrival got processed immediately) ,}{. That is, a state when a new arrival got processed immediately.}  \newrevise{The re-partitioning of the representative infinitesimal generator matrix followed by the relevant proofs morphed into formulations for $W(.)$ as the Laplance Transform $w(s)$ and the corresponding closed-form solution for the average time taken before being processed.}{From the formulations of the Laplance Transform, expressions for the stationary waiting time probabilities and a closed-form solution for the average waiting time were derived.} The authors additionally showed how application of randomization methods could yield for the characterization of the stationary waiting time probabilities $W(t)$ for \ac{QBD} processes. \newrevise{The proof assumed }{These definitions followed from the assumption }that, given a statistical initial state denoted as a vector $z$, there existed an $n^{th}$ transition step of the stochastic process such that the queue was idle (state $\theta$). 
The numerical analysis initialized a few system parameters to compute for some properties of rate matrix, the average queue occupancy $L$ \newrevise{and Little's Law as the basis for the arithmetic computation of other}{plus other system}  descriptors. 

\newrevise{As they attracted more recognition, matrix-geometric techniques to formulate solutions for steady state condition in a "join the shorter" queue \ac{M/M/C} setup with threshold jockeying permitted were also the subject of a documentation by \cite{Adan.Ivo.etal}.}{In contrast to earlier suggestions \cite{Gertsbakh1984TheSQ} \cite{11} that a solution for the steady state could not be achieved when the number of queues \textit{C} exceeded 2, \cite{Adan.Ivo.etal} disagreed to assert that the solution lay within the state transition space sub-division method used. In the \ac{M/M/C} setup, new entrants joined the shorter queue and jockeying was permitted given hitting a preset threshold defined around the difference between the queue sizes. It was argued therein that, although existing studies presumed that ergodic conditions could be derived from the sub-division of the state spaces, the solutions did not take the rate matrix into account. And that this necessitated revisiting the state space splitting approach. As a result therefore, the partitioning was based on sub-levels such that sets of sub-levels were mapped to sets of states.} 
The splitting was a factor of whether a sub-level's behavior was regular ($l = T, T+1, \ldots  $) or otherwise (${1, \ldots, T-1}$)\newrevise{, }{. Here, }\textit{l} denoted a collection of states and \textit{T} the jockeying threshold. \newrevise{The authors}{From these sub-divisions, it was} showed that, the condition for ergodicity was only possible when the system utilization $\rho<1$. 
After partitioning based on sub-levels, the generator matrix \textit{Q} took a different form that was irreducible but resolvable given \cite{neuts1994matrix}(1.7.1)'s theoretical findings. \newrevise{The stationary vectors were similarly split along categories with respect to the sub-levels to obtain a mapping from stationary probability vectors to sub-levels using the Eqs.~\eqref{eq:19} and \eqref{eq:20}.}{Following the sub-divisions of the stationary probability vectors, it was possible to generate mappings between the sub-levels and the probability vectors using Eqs.~\eqref{eq:19} and \eqref{eq:20}.}
\begin{equation}
    p{_l} = P{_T}R^{l-T},\revise{}{(}l < T)
    \label{eq:19}
\end{equation}
\begin{equation}
    D{_0} + RD{_1} + R{^2}D{_2} = 0
    \label{eq:20}
\end{equation}
\textit{where $D_0$,$ D{_1}$ and $D{_2} $ denoted square sub-matrices constituting states at all sub-levels $\ge T$ and $p{_l}$ or $p_{T}$ as stable probability vectors corresponding to the level $l$ (bound on $T$) that resulted from the split of the stable probability vector $p$}

\newrevise{Because one}{One} of the sub-matrices in the generator \textit{Q} took different dimensions due to the constituent states in the sub-level set\newrevise{, it was shown how given this difference in structure,}{. Then based on these structural deviations,} a solution for the rate matrix $R$ could be expressed. \newrevise{The evaluation derived its reference from suggestions by \cite{Ramswami1986AGC} that by ascertaining the maximum eigenvalue of rate matrix, one could easily express for the solution of R }{Borrowing from} \cite{Ramswami1986AGC}\newrevise{}{'s theoretical formulations for the maximum eigenvalue of the rate matrix, the solution for this matrix was defined } by Eq.~\eqref{eq:21}.
\begin{equation}
  R = \binom{0}{w}
 \label{eq:21}
\end{equation}
where \textit{$w = -v{(D_{1}+\rho^{c}D_{2})}^{-1} $, ($w=w_{0},....,w_{m-1}$), m being number of states at a given level that defined the dimension of the square sub-matrices. c was the number of queues available} 

Contrary to the mostly homogeneous setups studied, \cite{I.J.B.F} analyzed the behavior of two heterogeneous queues ($M/M/C=2$), each operating at a different service rate\newrevise{ to }{. The studies were purposely to apply matrix-geometric techniques to }derive expressions for sizes of the queues under stable conditions.  
New customers joined the shorter of either queues\newrevise{ (}{, }or were alternatively routed based on a probability of the sizes of the two queues were equal\newrevise{) and it }{. It} was allowed to switch from a longer to shorter one given some jockeying threshold $T$.    
The \newrevise{proof built on the authors' earlier findings}{work built on} \cite{AdanandWessels1,AdanandWessels2}'s studies, where it was shown that, in infinite time when no jockeying allowed between queues, the equilibrium probabilities ($p_{m,n}$, $m,n$ as queue sizes) of the queue sizes conformed to product-form solutions. \newrevise{Therefore, numerical evaluations under set parameter configurations were done to validate}{These observations were subject to verification on }  whether the same assertion was true when jockeying was permitted.
It was found that the evaluation held true only for a defined portion \textit{Q} ($max(m,n)>T$ and $(T,T)$) of states. Then observing the rates of change in state of the stochastic process for only this portion of states\newrevise{conclusively categorized the process as}{, it was conclusive that the process was} irreducible. This meant further analysis of this portion of states as a separate process (with distribution $q_{m,n}$) with a relation to the main process \newrevise{denoted by $p_{m,n}=q_{m,n}P(Q)$ (where $P(Q)$ was the probability that the portion $Q$ included the main process)}{(whose distribution was $p_{m,n}=q_{m,n}P(Q)$). Here $P(Q)$ was the probability that the portion $Q$ included the main process}. 
The product-form solution then derived from the notion of defining a set of metrics\newrevise{ that were a}{. These metrics were defined as a} factor of the arrival rate, the service rate of the shorter queue and the queue admission probability. Then, the general purpose principle was the basis for the derivation of equations that resolved for these metrics\newrevise{ and this}{. This} system of equations characterized for the steady-state and ergodicity conditions for this portion of states $Q$. 
The work sought to also draw comparisons between the product-form solution and the one ascertained using the matrix-geometric technique. \newrevise{}{The authors proved that when the size of the larger line exceeded the jockeying threshold $T$, because of the unique formation at the boundaries, the formulation of the solution was the product of the stationary probabilities of each service line. Hence that given the appropriate sub-division of the state space, the matrix-geometric method yielded the same solution.} For the geometric solution, the generator matrix emanated from sub-division of the steady-state probability vector into sub-levels\newrevise{ that}{. These sub-levels} grouped the states depending on the size $l$ of the longer queue ($l<T$ and $l\ge T$)\newrevise{ and associating each of these sub-levels }{. And each sub-level was associated }to a steady-state probability vector $\overrightarrow{p_{l}}$. \newrevise{Since}{It was noticeable that} some of sub-matrices ($A_{0}, A_{1}, A_{2}$) in the generator matrix $G$ were irreducible\newrevise{, using existing theorems for ergodicity conditions }{. Adopting theoretical foundations in \cite{neuts1994matrix} (Theorem 1.7.1, 1.7.11) was sufficient for the proof for ergodicity. That is, given knowledge about the maximum eigenvalue of the $R$ matrix, taking the unique structural properties of $A_{0}$,  $\overrightarrow{p}_{l}$ when $l>T$ could be computed for.} The the matrix-geometric formulation was found to bear similarities to the product-form solution. 

\newrevise{More work on a finite capacity $M/M/2$ queueing system was done by \cite{2} where, customers had the chance to change from one queue to another.}{ \cite{2}'s work then offered a performance comparison with slight adjustments in the $M/M/C=2$ system.}
The arrivals were simulated to imitate a Poisson distribution with rate $\lambda$, choosing the shorter of the two service lines\newrevise{ each operating with }{. Each operated at} exponentially distributed service rates ($\mu_{1}$ and $\mu_{2}$) and the capacity of each queue was restricted to $L$. If none of the service lines was shorter than the other, preference for one would follow probabilities ($\alpha$ or $\beta$, $\alpha+\beta=1$)\newrevise{while instantaneous}{. Instantaneous} jockeying was possible when one of the service line was empty. \newrevise{}{For the Markovian chain, the author defined a sequence of difference equations that characterized the probability ($P_{i,j}$) of queues (\textit{i,j}) being in equilibrium state. The compressed form of the difference equations ($A_{k-1}P_{k-1} + B_{k}P_{k} + C_{k+1}P_{k+1} = 0, k = 2,3,4,...., L-1 $ and $A_{L-1}P_{L-1} + B_{L}P_{L} = 0, k = L$) were first re-arranged to compose for a matrix block $A_{1}P_{1}+B_{2}P_{2}+C_{3}P_{3}=0$. Here $A, B$ and $C$ were sub-matrices that encapsulated the dynamics in queue sizes and traffic intensities. Then the evaluations for the sub-matrices were computed from the definition of column vectors that encoded state transition probabilities over all positions in both queues.} 
%
Theoretically, the proof required that the sub-matrix $B_{k}$ was invertible\newrevise{}{. This required computing the inverse of the sub-matrix and its determinants. Such that, there} existed an inverse of this sub-matrix only if its determinant did not evaluate to 0 for all values of the traffic intensity $\rho$.  
\newrevise{Based on $P^{+}_{k}$ being the theoretical solution for stability conditions,}{Then taking} $P^{+}_{k}$ as the theoretical solution for stability conditions, $ A_{L-1}P_{L-1} + B_{L}P_{L} = 0, k = L$ \newrevise{expressed for this solution as the}{was conditional for the existence of this determinant.} \newrevise{probability that the system was occupied to maximum capacity ($P_L=-R_{L}A_{L-1}P_{L-1},$ $k=L$) given that $R_{L}=B^{-1}_{L}$.}{This solution characterized for probability that the system was occupied to maximum capacity ($P_L=-R_{L}A_{L-1}P_{L-1},$ $k=L$) when $R_{L}=B^{-1}_{L}$ ($R$ being the rate matrix).} Iteration of computations for $R$ evolved into solutions for the differential equations expressed in terms of $P_{0,0}$ (probability that both queues were idle). The proof by induction on $k$ was essential to relate the queue intensities and the total number of customers in the system as $g_{k}=\rho^{k-2}g_{2}$ for $k=2,3,...,2L$\newrevise{ (given $g_{k}=Pr(N=k)$; $ N=N_{1}+N_{2}$ and \textit{$N_{i},$ $ i=1,2$ were the number of customers in either queue}) that formulated for the equilibrium probabilities as the M/M/2 system capacity was doubled (\textit{2L}) or as $L \implies \infty$}{. Based on this relation, equilibrium probabilities of the M/M/2 system were formulated for as the capacity doubled (\textit{2L}) or as $L \implies \infty$.} 
A numerical analysis \newrevise{included setting}{initialized} the arrival and service rates to different values\newrevise{ and }{. Then }comparisons were made with earlier results from the Conolly's model\cite{Conolly.B}. Performance evaluations for the effect of the system utilization on the equilibrium probability $g_{n}$ under different queue setups showed that the author's model was quantitatively better. 

\newrevise{Bulk workload migration received further attention in a  when  considered the variation in the sizes of the two service lines () hitting a preset threshold $L$ as the trigger for moving some workload $K(0<K<L)$ to the shorter queue. }{The more practical and challenging aspects arise when multiple tenants seeking to switch buffers in an instant. \cite{He2002TwoMQ} setup such a scene as two parallel $M/M/1$ servers. The portion of workload $K(0 < K < L)$ would be moved from the longer to the shorter of either queues ( $q_{1}$ to $q_{2}$ or vice versa) when the difference between their sizes hit the preset threshold $L$.} \newrevise{New Poisson distributed arrivals (at rate $\lambda_{i}$) joined service queues that run at exponentially distributed processing rates ($\mu_{i};$ $i=1,2$)}{Admissions obeyed a Poisson distribution with rate $\lambda_{i}$ while the processing times of either servers were exponentially distributed at rates $\mu_{1}$ and $\mu_{2}$ }. \newrevise{ to yield a}{This activity was generalized as} \ac{QBD} process $({(q_{1}(t), q_{2}(t)), t \ge 0})$ with state space ${(n_{1}, n_{2}) : n_{1} \ge 0},$ $| n_{1} -n_{2} | < L$ ( where $n_{i}, i=1,2$ were the number of customers in a given queue). \newrevise{}{According to the authors, the structure of the stochastic process made it hard to analyze. However, the process constituted special properties ($q_{1}-q_{2}\rlap{\kern.45em$|$}>L-1$) which, coupled with inherent recurrence properties of such processes, allowed for the sub-division of the state space $\{(q_{1}(t), (q_{2}(t)), t \ge0 \}$. This sub-division was essentially to convert the process into a valid \ac{QBD} to which matrix-geometric techniques could be applied for a solution to the equilibrium probabilities.} 
\newrevise{Then based on the inherent recurrence properties of such processes,}{ The studies then built on the theoretical propositions of \cite{JFCKingman,He2002TwoMQ}, that for such processes, steady-state conditions could only exist under specific conditions of the traffic intensity $(\rho<1)$.} 
This therefore meant re-organizing the state space until the resultant stochastic process ${(X(t),J(t), t \ge 0)}$ (individual queue occupancy as a level variable and the difference $J(t)=q_{1}-q_{2}$ in sizes as the phase variable) was irreducible and not dependent on the system occupancy $q(t)=q_{1}+q_{2}$. \newrevise{The equilibrium probabilities for this process were then formulated for using matrix-geometric methods based on earlier solutions ($A_{0}+RA_{1}+R^{2}A_{2}=0$) \cite{neuts1994matrix},\cite{GLatouche.VRamaswami} that entailed resolving for the eigenvalues (with the largest modulus - Perron-Frobesius eigenvalue) of the rate matrices $R$ at the different levels $(L)$.}{The matrix geometric solution was similar to \cite{neuts1994matrix},\cite{GLatouche.VRamaswami} and sought to resolve for the eigenvalues (one with the largest modulus - Perron-Frobesius eigenvalue) of the $R$ matrix at each of these different sub-division levels in $L$.} 
\newrevise{And these probabilities formed the basis for proof for stability conditions for system descriptors like likelihood that a queue was idle, number of serviced units, estimates on the number of consumers in a queue under steady conditions, how often workload was transferred $(T_{R,1->2}, T_{R,2->1})$ within the system, etc.}{Besides formulating for the stable conditions of common descriptors like system redundancy or number of processed tasks, the rate at which bulk load was moved around in the system was also expressed for.}
\newrevise{Expressions for the rate at which the distribution of system occupancy diminished were formulated and it was shown}{Expressions for the rate at which the distribution of system occupancy diminished were formulated. From this, it was deducted} that this decay rate was neither affected by the workload transfer threshold $L$ nor the number of workload migrations $K$ under varying inequalities of the traffic intensity $\rho$. 
The analytic solution was validated by numerical evaluations that involved experimentation with variations in queue design parameters. 
\newrevise{For example, the case of the first queue with $\lambda_{1}=1$, $\lambda_{2}=2$, $\mu_{1}+\mu_{2}=1=4$, $L=5, K=3$, it was observed that as the service rate increased, the rate at which customers were transferred to the second queue decreased and vice-versa. This}{The results} revealed evidence of the convexity properties characteristic of the relationship between the service rates and the rates at which customers were transferred from one queue to another. 
It was conclusively suggested that, choosing the right processing rate for each queue was requisite for even load distribution\newrevise{and that the}{. The} parameter that exerted much influence on these decisions was the deviation in queue utilization (traffic intensity). 

\paragraph{\revthree{\textbf{Limitations of Analytic Modeling in Next Generation Networks:} Matrix-geometric methods yield exact solutions for structured, stationary Markovian systems. However, emerging 5G/6G architectures that are characterized by slice heterogeneity, non-stationary dynamics, partial or delayed observations, migration/signaling costs, federated multi-vendor policies etc, invalidate the the assumptions in these analytic models.}
}


\revise{}{
\section{Behavioral Models}
}
\subsection{\revise{}{Modeling based on the Value of Information}}

\begin{figure} 
  \centering
  \includegraphics[width=0.53\columnwidth]{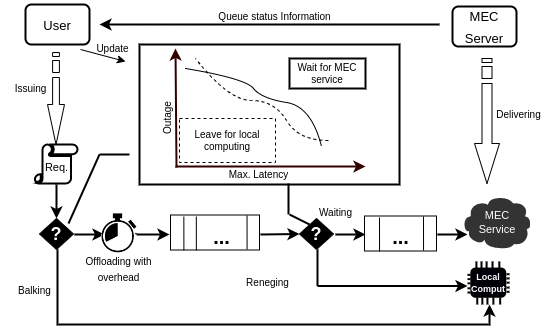} 
 \caption{\scriptsize Illustrating the impatient customer that had the option to either process a task on a \ac{MEC} server or locally depending on the latency requirements of the underlying application}
     \label{fig:offloading_tasks}
\end{figure}

Recent growth in latency-sensitive applications forces operators to prioritize traffic and manage selfish, impatient tenants. Han et al. \cite{7} modeled this impatience (reneging/balking, akin to jockeying) in an FCFS setting.
As depicted in Figure \ref{fig:offloading_tasks}, the customers had the option to continuously weigh the risk related to either process jobs locally (on their devices) or forward them to a cloud server. 
The preference for either local or cloud server was  a factor of the latency requirements of the jobs to be processed. These requirements were defined in terms of the expected cloud server response times $\tau_{c}$. Additionally, the predicted waiting time ($\tau_{w,k}$) of a user landing in position $k$ at joining time also influenced the decision for a processing platform. A reward $u$ that decays with total delay $\Delta t$ drives decisions: users maximize expected reward while minimizing risk $P_0$ from varying network conditions. With accurate channel information the model shows a user will not immediately withdraw a submitted task, since predicted remaining latency decreases as earlier jobs are served (see \cite{7} , Lemmas 1–2). The paper further studies decision behavior under imperfect information: learned experience and estimator error change perceived risk and can make users either overly cautious or rash. Numerical experiments quantified the regret, learning gain and reward or loss under correct versus partial system knowledge, highlighting how queue-state information quality shapes offload and jockeying choices.

\revise{}{\cite{pezoa}'s work then provided an extended analysis of the value of \newrevise{}{buffer status} information and its applicability in balancing the load within distributed compute systems. Here, the state of a node's knowledge about prevailing queue setup was fundamental for admitting tasks to the queues and transferring them around to alternative queues. The nodes cooperated on task executions by migrating excess load (total system load minus load on node) to other nodes. Equation~\eqref{eqn_partition} (the case of $n\geq 3$ for example) was definitive of the excess load partitioning and distribution to $n-1$ nodes. \newrevise{And}{The} transfer of this excess load depended on the size of the partition such that the least loaded node received the bigger partition. 
\begin{equation}
    p_{ij} =
    \begin{cases}       
        \frac{1}{n-2} \left( 1 - \frac{\lambda^{-1}_{{d}_{i}} Q_{i}(t-\eta_{{ji}})} {\sum_{l\neq j} \lambda^{-1}_{{d}_{l}} Q_{l}(t-\eta_{{jl}})} \right), \quad  \sum_{l\neq j} Q_{l}(t - \eta_{{jl}}) > 0 \\
        \frac{\lambda_{{d}_{i}}} {\sum_{k\neq j \lambda_{{d}_{k}}}},  \qquad \quad \qquad \qquad \quad   otherwise,   
    \end{cases}
    \label{eqn_partition}
\end{equation}
\textit{where $\eta_{{jl}}$ was the expected lag when node $l$ and node $j$ communicated, $\lambda_{d_{i}}$ was the rate at which a countable number of tasks departed the queue $i$. Also $\lambda_{d_{k}}$ denoted the departure rate at preset values of the $k^{th}$ load balancing instant. And $Q_{l}(t - \eta_{{jl}})$ was node $j$'s assumption about the number of tasks running on node $l$ which  depended on the communication lag not exceeding time $t$.}

\newrevise{ And the}{\noindent The} load balancing algorithm resident on each node was executed before accepting any collaboration in the task processing. \newrevise{The state of the knowledge local to the nodes required prior broadcasting of buffer sizes by all nodes within the cluster.}{The deployment of the algorithm followed from a prior broadcast from all nodes about their current buffer sizes as an update to the local state of the knowledge.}  Two load balancing policies were evaluated, i.e. centralized one-shot and dynamic load balancing. For the centralized one-shots rule-set, the proof evolved from adoption of principles of conditional expectation and regeneration-event decomposition\newrevise{ to express}{. From these theoretic findings, it was possible to characterize} for the average overall completion time (AOCT).
The centralized one-shots rule-set was extended for distributed environments as a sender-initiated dynamic load balancing (DLB) policy capable of meeting the dynamic processing speeds of the infrastructure.
For the one-shot centralized policy, \newrevise{experiments (two servers) to optimize the overall time to complete (AOCT) tasks}{experiments were setup with two servers with the aim of optimizing the overall time to complete (AOCT) tasks.} \newrevise{showed}{It was observed } that load balancing actions taken when $t_{b}$ increased beyond one second evolved into the slower node carrying more load. \newrevise{Hence the}{This resulted in } larger measures in the AOCT because of delayed updates to the knowledge state. 
The performance of the DLB policy was evaluated in terms of the time it took to complete a given task within a defined time frame. \newrevise{i.e. the mean task completion time ACTT as a combination of processing, queueing and transfer time .}{This metric, therein referred to as the mean task completion time (ACTT) was quantified from the processing, queueing and transfer time.} 
\newrevise{In conclusion,}{From the} analysis of the two qualitative measures (system processing rate (SPR) and ACTT) for both policies under different configurations of the load balancing gain $K$\newrevise{ led to the generalization}{, the following generalizations were drawn: In} either policies, improvements in SPR were recorded under lower measures in $K$\newrevise{}{. However, } more transfer activity in higher measures of $k\in K$ or excessive load migration delays that led to higher ACTT for the static load balancing policy (one-short)\newrevise{while the}{. The}  DLB policy on the other hand yielded lower queueing transfer delays to reduce the ACTT. \newrevise{Further benchmarking of the}{Further comparative assessment of the }proposed policies \newrevise{relative to}{against} classical DLB policies like Shortest-Expected-Delay (SED) or Never-Queue (NQ) revealed measurable improvements in the ACTT with the DLB over NQ and SED.
}

\revise{}{In a shift from centralized to decentralized control in multi-tenancy MEC environments, \cite{kiggundu2024resource} \newrevise{'s pioneering}{pioneered} work in behavioral modeling of impatience in queueing systems\newrevise{}{. Their decision model} sought to understand the benefits switching queues brings to the impatient tenant. \newrevise{It is argued that the state space curse in stochastic models of jockeying coupled with the centralized control of the behavior might not be practical given the dynamics inherent in next generation communication systems. That for decentralized management, }{The authors propose for decentralized control of the impatient tenant's behavior. This follows from the argument that current centralized control of this behavior might not be practical in next generation communication systems. That the inherent dynamics in such systems further explode the state space leading to intractable stochastic models. Hence, that the} rationale to move workload from one queue to another should be made by the individual tenants after assessing the up-to-date availed information about the expected waiting time. The Monte Carlo \newrevise{findings}{experiments} assumed a setup of network slices arranged as queues \newrevise{($M/M/C$, $C=2$), such that}{in an $M/M/C$, $C=2$ configuration. The} arrivals that obeyed a Poisson distribution with rate $\lambda$ and joined the shorter of the two heterogeneous buffer lines\newrevise{}{. This was assuming the new tenants were availed with }given prior knowledge about queue lengths. The tasks were processed at exponentially distributed service times. Then at each completion of a given task in either queue, tenants evaluated whether to stay in their current position(s) $k$ or to jockey to the alternative buffer line. \newrevise{The jockeying here disregarded whether jockeying was to the shorter queue but}{Here, there was disregard for whether jockeying was to the shorter queue or not. Instead,} the rational was premised on the expected waiting time that the jockeyed task would take. 
However, the position \newrevise{}{and therefore the effective expected waiting time} of the jockey in the alternate queue was a factor of what portion of the new arrivals $\beta \leq \lambda$ would prefer the same queue as the jockey\newrevise{ and to obscure}{. To manifest} this competitiveness therefore, the jockey’s final position followed a shuffle operation with the portion $\beta$ of new arrivals. \newrevise{The work assumed at least a single departure $N\geq 1$ occurred such that if $\beta \leq \lambda$ sought}{Assuming $N\geq 1$ departures at an instant and $\beta \leq \lambda$ arrivals seeking } to join the preferred queue at that point in time, then switching buffers was only if the expected waiting time in the preferred queue (at position $\tau$) was less than when the tenant stayed put (at position $k$). \eqref{eqn:arrival_choices} was definitive of \newrevise{}{this behavioral decision following from} these sojourn time-position dependencies.
\begin{equation}    
    F^{\overrightarrow{ij}}_{T_{w}| \tau}(t_{w}|\tau) =
    \begin{cases}       
       T_{w} \quad | \quad P^{\beta}_{Q_{i,j}}(t+1) \quad  & \text{if } \beta,N \geq 1\\
       T_{w} \quad | \quad P(N\geq 1) \qquad & \text{if } \beta = 0  \\
       T_{w} \qquad \qquad \qquad & \text{if } \beta, n = 0 
    \end{cases}
    \label{eqn:arrival_choices}
\end{equation}
\textit{where $T_{w}$ was the expected waiting time and $P^{\beta}_{Q_{i,j}}(t+1)$ denoted the probability that $\beta$ new arrivals joined the preferred queue at $t+1$ when the jockey decision was to be taken.} 

Formulations for the number of times that tenants switched from one queue to another then followed from the adoption of principles of conditional probability theory (Baye's theorem)\newrevise{to resolve for the dependencies between the expected waiting time, departures and new arrivals. It was shown therein that, from the Monte Carlo simulations, the sensitivity of the impatient tenant to perturbations in the system descriptors could first be assessed to extract correlations or dependencies that could guide the rationale to jockey. The results from the numerical evaluations of the model were a further revelation about the positive impact of switching buffers such that, the }{. The numerical analysis then compared the empirical measures from the Monte Carlo experiment to these analytical expression given the underlying dependencies. It was observed that} tenants that jockeyed more than once ended up waiting less until service in comparison to tenants that did not jockey at all.

Dispatching queue descriptor information as Markov models at varying intervals was investigated in \cite{kiggundu2025} to assess what kind of information or at what interval does this information about queue states lead to optimal system performance. The study involves disseminating two Markov models, one for the service rates of either queue and another Markov model of the how often the queue sizes change. The queued tenant then weighs the decision to jockey or renege based on the comparison of the Markov encapsulations of the service times and queue length dynamics for both queues. A tenant reneges from the queue to prefer the local processing if  the estimated remaining waiting time at a given position must be less than the time it takes to process the task locally given that there's no waiting involved. The jockeying tenant compares the statistical distributions of the service times using First-order Stochastic Dominance (FSD). A rule based policy is then devised such that the queue learns from the tenant impatience and adjusts it's service rates to minimize the delay and impatience. In their Monte-Carlo setup, the information models are disseminated at intervals of \textit{3,5,7,9} seconds and for each interval is characterized by hundreds of iterations. Numerical results reveal that indeed the interval at which broadcasting takes places has an effect on the eventual impatience of the tenants. The rule-based heuristic yields less volatility with regard to the impatience and sub-optimal service operations. 
}


\revise{}{
\subsection{Artificial Neural Networks modeling}
}

\newrevise{The application of jockeying}{Although not yet prevalent in jockeying modeling literature, active queue management techniques ideally embed \ac{AI} algorithms to optimize routing and workload distribution. One such study was} in \ac{FLP} problems \newrevise{ was investigated in}{ where} \cite{Bi0bjective} studied a cooperative multi-layered setup of queues. \newrevise{Jockeying had mostly been discouraged given the presumable associated costs and complexity that the behavior presented forthwith. }{}
The authors were interested in dynamically availing the facilities in each layer effectively so as to meet the emerging demand in small and large systems. \newrevise{}{The demand for service queues followed a Poisson distribution.} A job was processed through each layer by those facilities in closest proximity to the job's location\newrevise{ and t}{. The service rates at each facility were exponentially distributed and each facility in a given layer participated in the processing of a jockeyed applicant. T}he optimal solution \newrevise{was}{therefore involved} determining the set of facilities that could partake in the servicing of the job request.   
Starting with a solution for small-scale systems, expressions that characterized the objective functions (reduce jockeying plus mean waiting times and keeping all facilities busy) were formulated. 
\newrevise{The applicability of the augmented $\epsilon$-constraint method when evaluating for global solutions (Pareto-Optimal) to multi-objective non-linear models was restricted to small scale scenarios.}{For the small scale scenarios, an augmented $\epsilon$-constraint method was used to validate the global solution to the multi-objective model. } And for medium to large scale service systems, an \ac{NSGA}-II was deployed. \newrevise{}{The behavior of each entity in the population and system components like layers or facilities were abstracted as chromosomes. How worth a chromosome was for parenting the next generation was determined by associating that chromosome to penalties defined by $p(x)=U*Max\{0,\frac{g(x)}{b}-1\}$
\textit{(such that) $U$ was a constant, $g(x)$ the constraint and $p(x)$ the punishment on chromosome $x$}).}
The Taguchi scheme was found relevant for the adjustment of input parameters used for the initialization of the population. 
The final mating population evolved from the iterative application of conventional genetic operations (like selection, cross-over, etc) until the fitness function quantitatively yielded no better results for a series of selection runs.
The authors concluded the investigations with a numerical analysis of the model by setting up a manufacturing system\newrevise{and results were documented when jockeying was permitted from one facility to another within the the layers versus when applicants were processed by a single facility.  
The performance measurements}{. Performance comparisons were done when jockeying was permitted versus when no jockeying activity within the layers. Results further gravitated the benefits of jockeying not only at the system level but also at the consumer level. At the system level, the job transfer overhead was minimized plus the time queues were idle. This was because only facilities within the proximity of a potential jockey were those considered to participate in a job's sub-processing. Yet still at the user-level, the jockeying behavior was beneficial in terms of the overall time the user waits before being served. }

\paragraph{\revthree{\textbf{Limitations of Behavioral Modeling in Next Generation Networks:} These models assume timely, low-cost and accurate queue-state information and stationary service laws—assumptions that break down under mobility, NTN links, strict privacy constraints and heterogeneous, virtualized \ac{SDN} stacks.}
}

\section{\revise{}{Discussion, Conclusion and Future Work}}
\revthree{The dynamic migration of queued jobs between service pools has shown measurable performance improvements in MEC and heterogeneous 5G/Beyond environments \cite{han2023queue,Jianshan}. By leveling demand spikes, it raises average server utilization toward optimality. In the MEC, this is achieved by routing tasks across heterogeneous access points\cite{Jain,Mian, BV2022load}, slices and compute tiers to mitigate performance bottlenecks under variable traffic conditions \cite{Aloqaily,Schotten}. In end-to-end slice setups, task offloading that employs jockeying behavior reduces processing delays and improves service performance \cite{Lei,Poor}. Numerical studies show that adaptive workload redistribution reduces the mean-sojourn by up to $20-30\% $ to accelerate task completion for latency sensitive MEC applications \cite{DOWN2006509,pezoa,EVANS1993897}. The benefits extend to real-time load balancing use cases where jockeying heuristics reallocate jobs to minimize buffer growth and packet loss. When combined with network-level state information, they further improve application-specific resource allocation and utilization \cite{Lin,Tahir}. Additionally, under SLA regimes, jockeying can prioritize ultra low-latency traffic ~\cite{LWZ2022edge} and improve QoS/QoE when integrated with packet schedulers \cite{Zhang,Tom}, slice controllers or sub-band selection mechanisms \cite{abir_etal,Schotten,Shiang,mac_layer}. And reducing customer holding times or aggregate waiting minimizes service costs in energy constrained or SLA-penalized settings such as UAV assisted MEC \cite{7218459,Schotten}. }

\revthree{Fully realizing these qualitative benefits in next-generation systems demands careful attention to MEC and slice deployment constraints. That is, impatience manifested as jockeying introduces coordination overhead. The behavior can also destabilize scheduling given that it is sensitive to physical layer effects and mobility dynamics. Understanding and mitigating these factors is therefore essential before jockeying can move from a promising technique to a dependable production strategy in heterogeneous, multi-vendor networks. Besides understanding \textit{why} workload migration improves general system utilization, the previous sections in addition expose \textit{which} techniques and \textit{how} each model abstracts this practice.}

\revthree{However, two higher-level questions motivate reopening this line of inquiry for modern networks: (1) what architectural and technological changes are emerging in communication systems, and (2) are the classical modeling approaches still adequate under these changes, or do new issues arise that require different techniques? In response to these queries, Section~6.1 examines how these architectural trends in next-generation networks challenge traditional jockeying assumptions. We then identify concrete modeling and engineering challenges that arise from these architectural transformations.
}

\subsection{ \revthree{ Architectural transformations and the expected Operational challenges }}
\subsubsection{\revthree{Network Slicing Implications:}}
\revthree{ Third Generation Partnership Project ( 3GPP ) recommends a major evolution in how RAN and Core networks are structured. This change is called the functional split. Proposals hint decoupling the traditional Baseband Unit (BBU) and Remote Radio Unit (RRU) into two flexible components, that is, the Distributed Units (DU) and the Centralized Units (CU). These units have separate paths for the Control Plane and User Plane \cite{perez2023tutorial,Larsen}. The objective is to decompose the network functionality into modular software components and service chains that can be instantiated, scaled and composed by multiple vendors \cite{barakabitze20205g,ordonez2017network}.
These modular components, when combined into an end-to-end network slice expose a slice specific set of compute or network
resources, control policies and SLA metrics (latency, reliability, cost) \cite{elayoubi20195g,Foukas}. From a queueing theory perspective, each slice (or slice sub-component, e.g., an edge compute pool) can be modeled as a logical service
queue (or small queue network) with its own service rate characteristics and pricing. Tenants must then choose the best slice to minimize both cost and latency.
Because slices differ in both service parameters and cost schedules, this choice incentivizes impatience behaviors (reneging or jockeying) when slices become transiently adverse to the tenant’s objective\cite{Papageorgiou}. Allowing tenants to pick between vendor slices creates conditions where jockeying is common. This behavior has been found to help improve resource sharing between multiple vendors \cite{shu2020novel,wang2019enable}.
}

\subsubsection{ \revthree{Multi-vendor Challenges:}} 
\revthree{Unlike abstract queue networks, real MEC deployments must consider physical topologies and user mobility since MEC nodes span fronthaul and/or backhaul links. Hence, network delays and user movement complicates the optimal selection of target queues. Embedding jockeying policies within network slicing frameworks necessitates joint optimization of compute placement and communication cost. Moreover, designing jockeying-aware schedulers that guarantee latency and reliability SLAs across network slices is critical for future edge applications.
In such heterogeneous MEC environments, jockeying becomes a tool for dynamically shifting workloads between slices or queues to minimize perceived delay and operational expenses. However, to realize the full benefits in these environments, we find the following operational challenges that still have to be overcome.} 
\begin{itemize}    
    \item \revthree{ Real-time status dissemination: Broadcasting up-to-date queue lengths or time-of-flight estimates can congest control channels. While timely information aids routing \cite{Huang,Calasanz,Shiang_etal}, it adds overhead and may require lightweight, event-driven or piggybacked updates. Partial, inaccurate or maliciously altered status data could induce inefficient switching or mimic denial-of-service patterns.}
    \item \revthree{ Topology and mobility: Unlike abstract queue networks, MEC deployments must account for fronthaul and/or backhaul latency, user mobility, and multi-hop paths between collaborating nodes \cite{Maekawa,GAETA2006149}. These factors affect task transfer times and can undermine the timeliness of jockeying decisions.}
    \item \revthree{ Batch versus individual switching: Bulk migrations of queued jobs may improve throughput but risks transient overloads. Conversely, overly frequent switching at low jockeying thresholds can cause “ping-pong” behavior that wastes resources.}
\end{itemize}

\revthree{Addressing these issues demands cost-sensitive migration rules, authenticated state dissemination and context-aware policies that jointly optimize queue selection, compute placement, and communication cost under realistic network constraints. This necessitates quantifying the trade-offs between the performance gains from jockeying and the control or transfer overheads it incurs. The operational challenges identified above, in addition to migration and signaling costs, and cross-domain policy constraints, motivate a pragmatic reappraisal of design choices. In Section~6.2, we synthesize these challenges to guide suggestions for more practical hybrid architectures and propose mitigation techniques as part of the future work. 
}

\subsection{\revthree{Conclusion and Future Work}}
\revthree{Third Generation Partnership Project (3GPP) positions Quality of Experience (QoE) as a key metric for 5G and Beyond. In such networks where users expect predictable QoS or QoE, modeling selfish but impatient tenants is essential for slice admission and resource allocation policy design. Existing implementations ignore the need to separate state dissemination from decision control and therefore two design aspects must be distinguished: 
\begin{itemize}
    \item \textbf{Centralized dissemination}, involves aggregation and broadcasting of queue metrics (e.g., queue lengths, service rates, waiting times) from a central point to ensure system-wide visibility of queue states \cite{kiggundu2024resource,Shiang_etal,Antoine}. While centralized dissemination offers consistency, it can incur significant control-channel overhead. Conversely, centralized control without fresh visibility of system states risks delayed or suboptimal responses in dynamic MEC environments. 
    \item  \textbf{Centralized control}, in which a single authority enforces or overrides user decisions (forbids jockeying, enforces admission). Centralized control can be optimal when latency and freshness are guaranteed, but it suffers single point delays, scaling limits and fragility in large, mobile, or NTN deployments. Decentralized decision-making, where each user or local agent autonomously evaluates and acts upon available information, can still function effectively provided that dissemination of state information—centralized or otherwise—is sufficiently timely and accurate. Decentralized decision making scales better but depends critically on the timeliness and fidelity of disseminated state \cite{NEURIPS2021_99ef04eb}.
\end{itemize}
}

\revthree{A practical hybrid architecture separates the roles of publishing and decision making. Queue descriptor metrics (length, mean/or variance of wait,  tenant tier, signature etc) are collected by a telemetry aggregator and published through a topic-based publish/subscribe broker. Then the edge agents, such as MEC nodes, base-station proxies, or UE proxies, subscribe based on the QoS requirements. Centrally, the slice orchestrator retains slice-level enforcement (migration budgets, cool-downs, admission rules), while individual jockeying or reneging decisions are decentralized or remain local to meet latency constraints. 
To reduce the overhead, state could be piggybacked on existing control messages (e.g., Radio Resource Control /N2 signaling between the g-NodeB and control-plane function or lightweight periodic keepalives). Then the updates can be event-triggered such that they  are transmitted only when metrics cross meaningful thresholds.
This hybrid design minimizes decision latency and avoids single point performance bottlenecks. Centralized dissemination ensures global visibility, decentralized agents act quickly on local conditions, and the slice orchestrator can enforce global constraints. Evaluating such designs requires measuring decision latency, control plane throughput, and the trade-offs or performance gap compared to centralized optimal policies across diverse topologies and mobility patterns.}

\revthree{
We advocate for modeling impatience behavior through the Value-of-Information (VoI) approach. In this approach, high-fidelity updates are triggered only when the expected utility gain outweighs the communication cost \cite{Hassin.Haviv,zaiats}. A lightweight VoI surrogate can be used to map state, uncertainty, and action costs to the expected utility gain. Bayesian methods, such as ensembles or predictive intervals, provide estimates of information value and determine when refreshes or on-demand requests are necessary.
Predictors are expected to deliver both point estimates and calibrated uncertainty. For simple settings, lightweight Markov models are suitable, while in richer contexts neural network ensembles or Bayesian neural networks offer greater accuracy \cite{Liu2023,Junichi,Rahman}. The learning capacity of these neural models has motivated studies that move beyond statistical quantification of queueing descriptors. Instead, these data-driven methods increasingly make deductive inferences that support automation processes \cite{Rahman,Aussem_etal}. Their performance is qualitatively assessed using the \ac{RMSE}, calibration error, decision accuracy, and marginal QoE gains per signaling byte.
}

\revthree{Incorporating jockeying in next-generation networks requires integration with orchestration frameworks and strong safety controls. Migration decisions must account for service-level improvements relative to migration costs, with slice budgets, cooldown intervals, and fairness schemes to ensure stability \cite{Raz,D_Raz,Bensaou}. Telemetry and control channels need authenticated communication and safeguards against malicious reporting, while adversarial testing helps assess the resilience of these defenses. While these mechanisms establish the foundation for secure and policy compliant jockeying, the dynamics of unconstrained switching introduce new risks. Excessive jockeying can lead to unnecessary oscillations, wasted resources, and unstable scheduling. Stability can be improved with hysteresis thresholds, cool-down timers, explicit accounting of migration costs such as transfer delay and energy expenditure etc. Evaluating these safeguards requires measuring the number of switches per session, signaling reduction, migration overhead, and worst-case oscillations.
}

\revthree{There is need for empirical validation of the above proposals through practical investigations in multi-access edge computing (MEC) testbeds. These testbeds should embed telemetry brokers, MEC decision agents with predictors and value-of-information logic, orchestrator hooks, containerized migration, and radio emulation environments such as srsRAN or OpenAirInterface. And the testbeds should allow for controlled experiments that compare dissemination strategies (periodic, event-driven, or piggyback). The testbeds also need to incorporate mechanisms for assessing the role of value-of-information, uncertainty, and study performance under conditions of high mobility, flash crowds, or adversarial reporting. In addition, complementary large scale simulations, driven by real traffic traces, could help quantify migration latency, control plane overhead, energy costs, latency percentiles, task completion rates, and fairness indices. Then analytical studies, including bounds on minimum update frequency to maintain decision error within certain bounds or large-deviation asymptotics for rare events, provide further theoretical grounding. Together, these mechanisms constitute a key performance indicator suite (covering service quality, signaling overhead, computational footprint, and migration costs) that can guide robust deployment of jockeying in real-world networks.
}

\revthree{In conclusion, for systems where Quality of Service (QoS) and Quality of Experience (QoE) determine resource entitlement and pricing, selfish and impatient tenants pose a harder control problem. Their incentives to minimize completion time or maximize utility interact with slice-level policies and pricing, producing complex, feedback-driven behavior that simple rules cannot reliably manage.
This chronicle shows that classical jockeying models remain informative but are increasingly inadequate for 5G/6G deployments. Future models must explicitly represent slice heterogeneity and multi-vendor stacks, where non-uniform queue capacities, service disciplines, and admission rules invalidate homogeneous assumptions. They must also treat information flow as a constrained resource: latency, sampling rate, and signaling cost all affect whether a jockeying decision is actionable. Finally, models should include stability and security guarantees so controllers can limit oscillatory switching (ping-pong), bound migration overhead, and resist adversarial or corrupted state reports. Addressing these factors is necessary to produce robust, efficient, and audit-able impatience models suitable for research and system design in next-generation networks.
}

\begin{acks}
This work is supported in part by the German Federal Ministry of Education and Research (BMBF) within the project \textbf{Open6GHub} under grant numbers 16KISK003K and 16KISK004.
\end{acks}

\begin{acronym}[Queues]\itemsep0pt
\acro{ICT}{Information and Communications Technology}
\acro{QoS}{Quality of Service}
\acro{QoE}{Quality of Experience}
\acro{VoI}{Value of Information}
\acro{DTMC}{Discrete Time Markov Chain}
\acro{CTMC}{Continuous Time Markov Chain}
\acro{FCFS}{First -Come-First-Served}
\acro{LCFS}{Last-Come-First-Served}
\acro{QBD}{Quasi-Birth-Death}
\acro{3GPP}{Third Generation Partnership Project}
\acro{MEC}{Multi-access Edge Computing}
\acro{5G}{Fifth Generation}
\acro{6G}{Sixth Generation}
\acro{MAP}{Markov Additive Process}
\acro{SLA}{Service Level Agreement}
\acro{NSGA}{Non-Dominated Sorting Genetic Algorithm}
\acro{MOP}{multi-objective optimization problems}
\acro{MLCFLP}{Multi-Layered Congested Facility Location Problem}
\acro{FLP}{Facility location problems}
\acro{M/M/C}{Markovian/ Markovian/ number of queues}
\acro{G/G/C}{General/ General/ number of queues}
\acro{E/PH/C}{Erlang$(k)$/ Phase-type Distribution/ number of queues}
\acro{RAN}{Radio Access Network}
\acro{O-RAN}{Open Radio Access Network}
\acro{CP}{Control Plane}
\acro{DU}{Distributed Unit}
\acro{RRU}{Remote Radio Unit}
\acro{UP}{User Plane}
\acro{CU}{Control Unit}
\acro{SIRO}{Serve In Random Order}
\acro{IID}{ Identical and Independently Distributed}
\acro{BBU}{Baseband Unit}
\acro{SDN}{Software Defined Networks}
\acro{LTE}{Long Terminal Evolution}
\acro{ORAN}{Open Radio Access Network}
\acro{MEC}{Multi-Access Edge Computing}
\acro{SDN}{Software Defined Network}
\acro{NSSF}{Network Slice Selection Function}
\acro{SlaaS}{Slice-as-a-Service}
\acro{HAPS}{High Altitude Platform Station}
\acro{UAVs}{Unmanned Aerial Vehicles}
\acro{LEO}{Low Earth orbit}
\acro{AI}{Artificial Intelligence}
\acro{ML}{Machine Learning}
\acro{TLS}{Transaction Layer Security}
\acro{NTN}{Non Terrestrial Networks}
\acro{AQM}{Active Queue Management}
\acro{NFV}{Network Functions Virtualization}
\acro{MDPs}{Markov Decision Processes}
\acro{POMDPs}{Partially Observable Markov Decision Processes}
\acro{VoI}{Value of Information}
\acro{Dec-POMDPs}{Decentralized Partially Observable Markov Decision Processes}
\acro{RMSE}{Root Mean Square Error}
\end{acronym}

\bibliographystyle{ACM-Reference-Format}
\bibliography{jockey.bib}
\end{document}